\documentclass[12pt]{iopart}

\usepackage{iopams}

\usepackage{graphicx}
\usepackage{epsfig}

\usepackage{amsfonts}
\usepackage{amssymb}
\usepackage{mathptmx}      

\usepackage{color}

\usepackage{psfrag}

\newcommand{\mc}{\mathcal}
\newcommand{\lb}{\label}

\begin{document}

\title[U(1) LGT Connects All Classical Models with Continuous Variables]{The U(1) Lattice Gauge Theory Universally Connects All Classical Models with Continuous Variables, Including Background Gravity}

\author{Ying Xu$^{1,2}$, Gemma De las Cuevas$^{1,2}$, Wolfgang D\"ur$^{2}$, Hans J. Briegel$^{1,2}$ and Miguel A. Martin--Delgado$^{3}$}

\address{$^{1}$ Institut f\"ur Quantenoptik und Quanteninformation,\\ Technikerstra\ss e 21a, 6020 Innsbruck, Austria}
\address{$^{2}$ Institut f{\"u}r Theoretische Physik, Universit{\"a}t Innsbruck,\\ Technikerstra\ss e 25, A-6020 Innsbruck, Austria}
\address{$^{3}$ Departamento de F\'{\i}sica Te\'orica I, Universidad Complutense, 28040 Madrid, Spain}

\ead{\mailto{ying.xu@uibk.ac.at}, \mailto{gemma.delascuevas@uibk.ac.at}, \mailto{wolfgang.duer@uibk.ac.at}, \mailto{hans.briegel@uibk.ac.at}, \mailto{mardel@miranda.fis.ucm.es}}

\begin{abstract}
We show that the partition function of many classical models with continuous degrees of freedom, \textit{e.g.} abelian lattice gauge theories and statistical mechanical models, can be written as the partition function of an (enlarged) four--dimensional lattice gauge theory (LGT) with gauge group $U(1)$. This result is very general that it includes models in different dimensions with different symmetries. In particular, we show that a $U(1)$ LGT defined in a curved spacetime can be mapped to a $U(1)$ LGT with a flat background metric. The result is achieved by expressing the $U(1)$ LGT partition function as an inner product between two quantum states.
\end{abstract}

\pacs{03.67.-a, 11.15.Ha, 03.67.Lx, 75.10.Hk, 05.50.+q}
\vspace{2pc}
\noindent{\it Keywords}: Lattice gauge theory, Statistical mechanics, Quantum information
\maketitle

\section{Introduction}
\label{intro}

The partition function $\mc{Z}$ is the keystone in both statistical mechanics \cite{Pa} and quantum field theory \cite{Cr83}. Thermodynamic quantities, like the free energy, the entropy, correlation functions, \textit{etc.} can be evaluated once $\mc{Z}$ is known as a function of physical parameters (such as the inverse temperature $\beta$, couplings constants $J$, external fields and other order parameters). On the other hand, gauge theories have proven to be crucial in the description of nature.
In particular, quantum electrodynamics (QED) is described by a $U(1)$ gauge theory.

It has recently been shown \cite{DDBM1,DDBM2} that the partition function of any classical spin model can be mapped to that of an (enlarged) four--dimensional ($4D$) lattice gauge theory with gauge group $\mathbb{Z}_2$ (see also the original idea \cite{VDB2} and closely related works \cite{DDVB, Va09, De10b}). More precisely, if one tunes the coupling strengths of the partition function of a (large enough) $4D$ $\mathbb{Z}_2$ LGT, this would equal the partition function of a classical spin model in any dimension, with any type of interaction pattern (including arbitrary many--body interactions), and thus also includes models with local and global symmetries. To obtain this, one expresses the partition function of a large class of models as an inner product between two quantum states, and then relates the quantum states (see \cite{HVDB} for a detailed treatment of the quantum formulation). The result is very general and unifies very different models. In this sense, the $4D$ $\mathbb{Z}_2$ LGT is a \textit{complete} model for this class of discrete models.
In fact, these mappings from a general discrete model to a $\mathbb{Z}_2$ LGT allow one to gain structural insight.

In the present work, we show that the partition function of many continuous classical models can be expressed approximately (to arbitrary precision) as the partition function of the $4D$ $U(1)$ LGT. The result holds exactly for a large class of models,
namely, models whose Hamiltonian has a finite Fourier series and no constraints on the variables. This includes models of different geometry, and in arbitrary dimensions ($1D$, $2D$, $3D$, \textit{etc.})
In the proof of the statement we generalize the quantum formulation developed previously \cite{DDVB,HVDB,VDB1,VDB2} and generate the (truncated) Fourier series of any target Hamiltonian. In parallel with the $4D$ $\mathbb{Z}_2$ LGT as the \textit{complete} model for discrete models, the $4D$ $U(1)$ LGT in this sense is \textit{complete} for a certain class of continuous models. We also define and consider the $U(1)$ LGT defined in a curved spacetime background. We show that this class of models are also included in our completeness result, \textit{i.e.} they can also be mapped to the flat $4D$ $U(1)$ LGT. The physical content of this result is the following: as long as the gravity coupled to the QED is a fixed background, it can be absorbed as if it were a flat spacetime as far as completeness is concerned. This situation has a physical counterpart in cosmology where one finds photons in an approximate fixed curved spacetime.

In this paper we will first present some basics on the $U(1)$ LGT in $\S\,\ref{basics}\,$. Then we will prove the completeness of the $U(1)$ LGT in $\S\,\ref{complete}\,$. Further illustrations of the completeness result with some examples and applications will be given in $\S\,\ref{examples}\,$. We will proceed to the $U(1)$ LGT in a curved spacetime and relate to the main result in $\S\,\ref{lgtwithg}\,$. In $\S\,\ref{generalize}\,$ we will generalize our result to a larger class of models. Finally, the conclusions will be drawn in $\S\,\ref{conclusion}\,$.

\section{Basics on the U(1) Lattice Gauge Theory}
\label{basics}

LGTs are gauge theories on a lattice representing a discrete spacetime. Generically, LGTs are useful non--perturbative formulations of gauge theories which allow for numerical simulations, \textit{e.g.} using the Monte Carlo methods \cite{Pan05}. This allows one to go beyond perturbative calculations with Feynman diagrams. The abelian $U(1)$ LGT was introduced by Wilson \cite{Wi74} and Polyakov \cite{Po75a,Po75b} as a generalization of Wegner's Ising gauge theories \cite{We71}.
The $U(1)$ LGT can be considered a discretization of electrodynamics (a pure gauge theory with an abelian gauge group $U(1)$) defined on discrete spacetime~\cite{Ko79}.
Non--abelian continuous LGTs are successful as having been proven to be asymptotically free \cite{Gr73} in the weak coupling limit, and important in the study of quark confinement at the strong coupling limit \cite{Wi74}.

For abelian LGTs it is possible to simulate numerically the continuum limit, $\Delta\to 0$, in an appropriate way, and verify that the resulting theory is the actual QED without confinement. This verification is important since on the lattice, most LGTs show confinement which is an artifact. It so happens that in the strong coupling limit of an LGT, the property that the gauge group is compact is essential for observing confinement, regardless of whether it is abelian or non--abelian. Related to this, the phase diagram of LGTs is very rich and relevant for taking the continuum limit \cite{Ko79}, and the $U(1)$ LGT serves as a test ground for this purpose. Unlike the non--abelian $SU(N)$ LGTs, which have a discrete center subgroup $Z_N$, the abelian $U(1)$ LGT has a continuous center subgroup which is identical to the group itself. The role of the $U(1)$ group on the confinement to deconfinement transition is also of interest from the point of view of the abelian dominance hypothesis for confinement which holds that a $U(1)$ subgroup controls the non--perturbative dynamics of non--abelian gauge theories \cite{tHooft}.

In the mean time LGTs have emerged as interesting theories by themselves. They are examples of models with local symmetries and non--local order parameters. They exhibit phases which do not appear in the continuum limit \cite{Ko79}. The phases of the $U(1)$ LGT can be characterized by the Wilson loop (see (\ref{wilsonloop}) below) which is an gauge invariant order parameter of the model. In the confined phase, this order parameter obeys an area law, whereas in the unconfined phase it obeys a perimeter law. The 't Hooft loop is a dual variable to the Wilson loop, and it constitutes another order parameter.

There are compact and non--compact $4D$ $U(1)$ LGTs \cite{Pan05, CR,Palumbo,PSS}. In the compact $U(1)$ LGT the degrees of freedom are exponentials of the edge degrees of freedom. Thus, they are directly elements of the $U(1)$ group which is compact, thereby the name. On the other hand in the non--compact $U(1)$ LGT the degrees of freedom are associated directly to the edges of the lattice, which do not need to be elements of the $U(1)$ group.
The former one only has a deconfined phase with massless photons, thus one recovers QED in the limit of continuous spacetime.
The latter one has two phases: a weak coupling phase, where the model has massless photons (gapless excitations), and a strong coupling phase, where there are massive photons and magnetic monopoles. The photons are screened by the monopoles, which corresponds to a mechanism of confinement of electrical charge. These phases are characterized with Wilson loops and 't Hooft loops as mentioned above.
In this paper, we will deal with the compact $U(1)$ LGT.

\begin{figure}[htb]
	\centering
		\includegraphics[width=0.6\columnwidth]{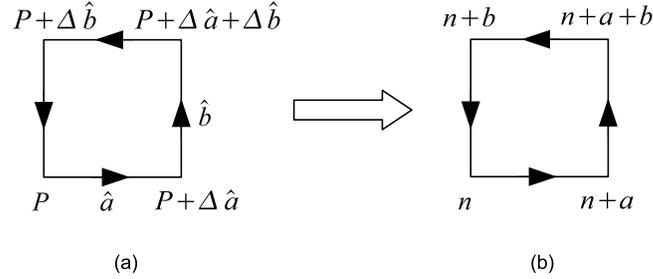}
	\caption{(a) A face of the square lattice. $\hat{a}$ and $\hat{b}$ are unit basis vectors and each edge is assigned a direction $\pm\hat{a}$ or $\pm\hat{b}$. The four vertices of the face are labeled by $P$, $P+\hat{a}\Delta$, $P+\hat{b}\Delta$ and $P+\hat{a}\Delta+\hat{b}\Delta$. (b) For notational convenience, we 
	use a single labeling for the vertices, namely, $n$ (the $n^{\mathrm{th}}$ vertex), $n+a$, $n+b$ and $n+a+b$, with $a\equiv\hat{a}\Delta$ and $b\equiv\hat{b}\Delta$.}
	\label{fig:face}
\end{figure}

\begin{figure}[htb]
	\centering
		\includegraphics[width=0.8\columnwidth]{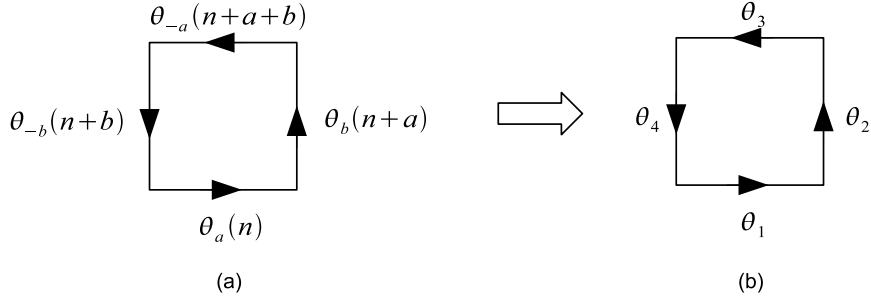}
	\caption{(a) The gauge fields $\theta_{a}$ as defined along the edges. (b) A further simplified notation where the fields along the four edges of a face are denoted by $\theta_{1}$, $\theta_{2}$, $\theta_{3}$ and $\theta_{4}$, and $\theta_{e}$ in general.}
	\label{fig:field}
\end{figure}

We briefly summarize the formulation of the $U(1)$ LGT as follows (with our notation close to that of \cite{Ko79}). It is illustrative to derive the Lagrangian of classical electrodynamics from that of $U(1)$ LGT in the limit of continuous spacetime~\cite{Ko79}. Let us consider a $4D$ square lattice as the discretized spacetime. Let $\Delta$ denote the lattice spacing and $\hat{a}$, $\hat{b}$ \textit{etc.} the unit basis vectors. We denote vertices, edges and faces of the lattice by $v$, $e$ and $f$, respectively. The set of all vertices, edges and faces is denoted by $V$, $E$, and $F$ and the number of elements in each set by $|V|$, $|E|$, and $|F|$ accordingly. We choose a direction for each edge, see Fig. \ref{fig:face}.
The gauge field $A_{a}(n)$ is defined along each edge incident to vertex $n$ with direction $\hat{a}$ indicated by the sub--index (see Fig.~\ref{fig:field}). The Wilson loop for an elementary face (such as in Fig. \ref{fig:face}(a)) is
\begin{eqnarray}
	 U_{\mathrm{face}}&=&\mathrm{e}^{\mathrm{i}\Delta\mathcal{G}A_{a}(P)}\mathrm{e}^{\mathrm{i}\Delta\mathcal{G}A_{b}(P+\Delta\hat{a})}\mathrm{e}^{-\mathrm{i}\Delta\mathcal{G}A_{a}(P+\Delta\hat{b})}\mathrm{e}^{-\mathrm{i}\Delta\mathcal{G}A_{b}(P)}\nonumber\\
&=&\mathrm{e}^{\mathrm{i}\Delta\mathcal{G}\left([A_{b}(P+\Delta\hat{a})-A_{b}(P)]-[A_{a}(P+\Delta\hat{b})-A_{a}(P)]\right)}, \label{wilsonloop}
\end{eqnarray}
in which $\mathcal{G}$ is the coupling constant. The exponential of the second line in (\ref{wilsonloop}) resembles the field tensor $F_{ab}=\partial_{a}A_{b}-\partial_{b}A_{a}$. The action is constructed from the Wilson loops. for notational convenience, we switch to the simplified notation as in Fig. \ref{fig:face}(b).
We make the following redefinition of the gauge field:
\begin{eqnarray}
\theta_{a}(n)&\equiv&\Delta\mathcal{G}A_{a}(P),\label{redef}
\end{eqnarray}
as well as the following conventions:
\begin{eqnarray}
\theta_{-a}(n+a)&\equiv&-\theta_{a}(n),\nonumber\\
\Theta_{ab}(n)
&\equiv&\theta_{a}(n)+\theta_{b}(n+a)+\theta_{-a}(n+a+b)+\theta_{-b}(n+b).\label{redefinition}
\end{eqnarray}
We can take the redefined gauge field $\theta_{a}$ as defined on the edges instead of on the vertices, see Fig. \ref{fig:field}. With these redefined gauge field notations, the exponential in (\ref{wilsonloop}) is equal to $\mathrm{i}\Theta_{ab}$ (which is a sum over the four $\theta_{e}$ field variables along the edges of a face, $e\in f$). $\Theta_{ab}$ can be thought of as defined for each face and denoted alternatively by $\Theta_{f}$:
\begin{eqnarray}
	\Theta_{f}=\theta_{1}+\theta_{2}-\theta_{3}-\theta_{4}=\sum_{e\in \partial f}\chi_{e}\theta_{e}. \label{Theta}
\end{eqnarray}
where $\partial f$ denotes the boundary of face $f$. The signs of the four edge variables in (\ref{Theta}) are chosen to be consistent with Fig. \ref{fig:field} (b). In general, a face can contain edges pointing clockwise and others counterclockwise. So that $\chi_{e}=1[-1]$ if $e$ is oriented clockwise[counter--clockwise]. (Note that only the relative sign of the $\theta_e$'s is relevant for the action (\ref{action}) below.) Now the Wilson--Kogut action of the $U(1)$ LGT can be written as
\begin{eqnarray}
	S=\frac{1}{2\mathcal{G}^{2}}\sum_{f\in F}\left[1-\cos{\Theta_{f}}\right]. \label{action}
\end{eqnarray}
This (Euclidean) action in the naive continuum limit is seen to be $\frac{1}{4\mathcal{G}^{2}}\sum_{f\in F}\Theta^{2}_{f}\sim \frac{1}{4}\int\mathrm{d}^{4}x\ F^{2}_{ab}$, consistent with the classical $U(1)$ theory. Also, the action takes the same form as in (\ref{action}) in any spacetime dimension.
This action is invariant under the gauge rotation applied on any vertex $v$:
\begin{eqnarray}
g_{v} = \prod_{e:\, v\in\partial e} U_{e}^{\chi_{e}}, \qquad e \in E \label{gaugesymm}
\end{eqnarray}
where $U_{e}$ is an element of the gauge group, in our case, $U_{e}\in U(1)$, and $\partial e$ denotes the boundary of $e$ (\textit{i.e.} the product in (\ref{gaugesymm}) applies to all edges $e$ incident to vertex $v$). Finally, the partition function (which corresponds to the Euclidean path integral) of this pure $U(1)$ gauge theory on a lattice takes the following form:
\begin{eqnarray}
\mathcal{Z}=\int^{\pi}_{-\pi}\left(\prod_{e\in E}\mathrm{d}\theta_{e}\right)\exp\left\{-\frac{1}{2\mathcal{G}^{2}}\sum_{f\in F}\left[1-\cos{\Theta_{f}}\right]\right\}. \label{partition}
\end{eqnarray}
Note that the first constant term in the sum of the exponential in (\ref{partition}) only introduces an overall constant factor which can be omitted.\footnote{Ignoring constant factors of the partition function corresponds to an overall shift in the action (or the Hamiltonian) which never changes any observable nor the equations of motion. It only introduces an additional constant to the free energy as we take the logarithm. Observables only relate to differences (or derivatives) in free energy.} In addition, the coupling constant $\mathcal{G}$ can be made face--dependent. Accordingly, we will start with the partition function of the following inhomogeneous model with local coupling constant $J_{f}$ for each face (The coupling constant $\mathcal{G}$ and (possibly) an inverse temperature $\beta$ are all contained in the local $J_{f}$'s):
\begin{eqnarray}
\mathcal{Z}=\int^{\pi}_{-\pi}\left(\prod_{e\in E}\mathrm{d}\theta_{e}\right)\exp\left\{\sum_{f\in F}J_{f}\cos{\Theta_{f}}\right\}.
\label{start}
\end{eqnarray}

\section{4D U(1) Lattice Gauge Theory as a Complete Model}
\label{complete}

In this section we show that the $4D$ $U(1)$ LGT is complete in the sense that a large class of classical partition functions with continuous degrees of freedom can be represented as special cases of the partition function of this model. In $\S\,\ref{hamiltonian}\,$ we define the set of models considered in our completeness result. Then, we present some tools in $\S\,\ref{quantum}\,$ and $\S\,\ref{ssec:rules}\,$ that we will require for the proof of the main result.
Finally, our completeness result is proved in $\S\,\ref{completeness}\,$.

\subsection{Class of Classical Models}
\label{hamiltonian}

Our completeness results embrace all continuous classical `spin' models, \textit{i.e.} models with a Hamiltonian satisfying the following conditions:
\begin{enumerate}
	\item The Hamiltonian depends on a set of $N$ continuous real variables $\{x_{j}|j=1, 2, \ldots, N\}$ and each variable takes value within a finite interval, \textit{i.e.}
	\begin{eqnarray}
	x_{j}\in [\,a_{j}\,,\, b_{j}\,], \qquad \forall j. \label{condition1}
	\end{eqnarray}
	\item The Hamiltonian is a sum over $K$-body interactions with $1\leq K\leq N$, \textit{i.e.}
	\begin{eqnarray}
	\mathcal{H}\left(\{x_{j}\}\right)=\sum^{N}_{K=1}\sum_{\{\mathrm{K-body}\}} H^{(K)}\left(\{x_{j}\}\right). \label{condition2}
	\end{eqnarray}
	\item There is no constraint on this set of variables, \textit{i.e.}
	\begin{eqnarray}
	x_{j} \quad \mbox{with} \quad j=1,2,\ldots,N \quad \mbox{are independent variables}. \label{condition3}
	\end{eqnarray}
\end{enumerate}
Furthermore, we assume that each $K$-body Hamiltonian is a well--behaved function which allows a Fourier series expansion over the $x_{j}$ variables. (Note that the spacial dimension of models under consideration is arbitrary.)

With the above conditions, we first normalize the ranges of variables by a linear change of variables from $\{x_j\}$ to $\{\theta_j\}$
\begin{eqnarray}
	\theta_{j}=\frac{2\pi}{b_{j}-a_{j}}\left(x_{j}-\frac{a_{j}+b_{j}}{2}\right), \qquad \forall j \label{thetaj}
\end{eqnarray}
such that each $\theta_{j}\in [-\pi, \pi]$. The relation (\ref{thetaj}) can be inverted $x_{j}=\frac{b_{j}-a_{j}}{2\pi}\theta_{j}+\frac{a_{j}+b_{j}}{2}$ so that the Hamiltonian is reexpressed in terms of these normalized variables and denoted by $\mathcal{H}\left(\{\theta_{j}\}\right)$. Next, we make a Fourier series expansion of each $K$-body Hamiltonian $H^{(K)}\left(\{\theta_{j}\}\right)$ over the $\theta_{j}$ variables:
\begin{eqnarray}
	 H^{(K)}\left(\{\theta_{j}\}\right)&=&\sum_{\{m_{j}\}}\mathrm{H}^{(K)}_{\{m_{j}\}}\exp\left(\mathrm{i}\sum^{\mbox{\tiny{K terms }}}_{j}m_{j}\theta_{j}\right)\nonumber\\
	&=&\sum_{\{m_{j}\}}\Re\mathrm{H}^{(K)}_{\{m_{j}\}}\cos\left(\sum^{\mbox{\tiny{K terms }}}_{j}m_{j}\theta_{j}\right)-\Im\mathrm{H}^{(K)}_{\{m_{j}\}}\sin\left(\sum^{\mbox{\tiny{K terms }}}_{j}m_{j}\theta_{j}\right). \label{fourier}
\end{eqnarray}
where $m_{j}\in \mathbb{Z}$, $\forall j$ and the Fourier coefficients $\Re\mathrm{H}^{(K)}_{\{m_{j}\}}$ and $-\Im\mathrm{H}^{(K)}_{\{m_{j}\}}$  are all real ($\Re$ and $\Im$ denotes real and imaginary parts, respectively). Therefore the general Hamiltonian $\mathcal{H}$ in (\ref{condition2}) is now written as a Fourier series over a set of basis functions
\begin{eqnarray}
	\{\cos\left(\sum_{j}m_{j}\theta_{j}\right), \sin\left(\sum_{j}m_{j}\theta_{j}\right)\big|\, |m_{j}|=0, 1, 2,\ldots,\ j=1,\ldots,N\}. \label{basis}
\end{eqnarray}

\subsection{Quantum Formulation}
\label{quantum}

We present a quantum representation of the partition function $\mathcal{Z}$ in (\ref{start}). First, assign a quantum state $|\Theta_{f}\rangle$ to each face:
\begin{eqnarray}
|\Theta_{f}\rangle
=|\sum_{e\in f}\chi_{e}\theta_{e}\rangle. \label{thetastate}
\end{eqnarray}
Then we define the following quantum state $|\psi\rangle$ which contains the interaction pattern (By `interaction pattern' we mean the lattice or graph structure representing which variables are interacting):
\begin{eqnarray}
|\psi\rangle=\int^{\pi}_{-\pi}\left(\prod_{e\in E}\mathrm{d}\theta_{e}\right)\bigotimes_{f\in F}|\Theta_{f}\rangle. \label{psi}
\end{eqnarray}
Next we define another state $|\alpha\rangle$ which contains the (Euclidean) weight for each configuration of the $\theta_{e}$ variables of the whole lattice:
\begin{eqnarray}
	|\alpha\rangle=\int^{\pi}_{-\pi}
	\left(\prod_{e\in E}\mathrm{d}\theta_{e}\right)
	 \left(\prod_{f\in F}\mathrm{e}^{J_{f}\cos{\Theta_{f}}}\right)
	 \bigotimes_{f\in F}|\Theta_{f}\rangle.
\label{alpha}
\end{eqnarray}

In the following we show that $\langle\alpha|\psi\rangle$ is proportional to the partition function $\mathcal{Z}$ in (\ref{start}). Let us consider the linear transformation from the edge variables (gauge fields) $\theta_{e}$ to the face variables (field tensors) $\Theta_{f}$. We define two column vectors:
\begin{eqnarray}
 \boldsymbol{\theta}&=&
 \left( \theta_{1}  \ldots \theta_{e}  \ldots \theta_{|E|}\right)^t\,  \nonumber\\
 \boldsymbol{\Theta}&=&
  \left( \theta_{1}  \ldots \theta_{f}  \ldots \theta_{|F|}\right)^t\, .
 \label{columns}
\end{eqnarray}
where $^t$ denotes transposition. They are related by a linear transformation $\boldsymbol{\Theta}=\mathbf{I}\cdot\boldsymbol{\theta}$
where $\mathbf{I}$ is the face--edge incidence matrix with matrix elements $I_{f,e}=1$ if $e\in \partial f$ and $0$ otherwise.

We always assume that all edge variables are independent. \textit{i.e.} $\mathrm{rank}\mathbf{I}=|E|$. For a square lattice with periodic boundary conditions in $D$-dimension ($D\geq2$), there is a relation between $|E|$ and $|F|$ such that $|F|=\frac{D-1}{2}|E|$. If $|F|\leq |E|$ we add a set of linearly independent auxiliary face variables $F_{\mathrm{aux}}$, where $|F_{\mathrm{aux}}|=|E|-|F|$. This results in an $|E|\times |E|$ incidence matrix $\tilde{\mathbf{I}}$, \textit{viz.}
\begin{eqnarray}
 \boldsymbol{\Theta}=
 \left( \Theta_{1}  \ldots  \Theta_{|F|}\ldots\Theta_{|E|}  \right)^t=
\tilde{ \mathbf{I}}\cdot\boldsymbol{\theta}\, .
\end{eqnarray}
Geometrically, this amounts to saying that we are introducing new auxiliary faces to the original lattice and these new faces need not be squares.
If $|F|\geq |E|$, we pick up a maximally independent set of face variables and eliminate linearly dependent rows until the incidence matrix has a size $|E|\times|E|$. So that we may write $\boldsymbol{\Theta}= \left( \Theta_{1}  \ldots  \Theta_{|E|} \right)^t=\tilde{ \mathbf{I}}\cdot\boldsymbol{\theta}$, with again an $|E|\times|E|$ incidence matrix $\tilde{\mathbf{I}}$. Because not all face variables appearing in the quantum states $|\psi\rangle$ (\ref{psi}) and $|\alpha\rangle$ (\ref{alpha}) are independent, when we take the inner product of $|\psi\rangle$ with $|\alpha\rangle$, formal infinities would arise. These infinities, \textit{e.g.} $\delta^{2}(\Theta_{f}-\Theta'_{f})=\delta(\Theta_{f}-\Theta'_{f})\delta(0)$, always take the form of powers of $\delta(0)$. This problem can be treated in the following way. We introduce a large `momentum' truncation $\Lambda$ in the Fourier space (inverse space) for each face variable. So that $\delta(\Theta=0)=\frac{1}{2\pi}\sum_{n\in\mathbb{Z}}\mathrm{e}^{\mathrm{i}n\Theta}\sim\frac{\Lambda}{2\pi}$ with $\Lambda\to\infty$. Therefore when taking the inner product we shall have an overall factor $\left(\frac{\Lambda}{2\pi}\right)^{|F|-|E|}$ which tends to infinity formally. However, this factor does not affect any observable because it only introduces an additive constant to the free energy as we take the logarithm of the partition function. We shall call this a `regularization' method that allows us to make sense of the expressions that come out with harmless infinities. (A similar constant factor has been omitted going from (\ref{partition}) to (\ref{start}), see also (\ref{addfactor}) below and the argument there for comparison.)
Finally we see that $\langle\alpha|\psi\rangle$ is proportional to the partition function $\mathcal{Z}$ in (\ref{start}):
\begin{eqnarray}
&&\langle\alpha|\psi\rangle\nonumber\\
	&=&\int^{\pi}_{-\pi}\left(\prod_{e\in E}\mathrm{d}\theta^{\prime}_{e}\mathrm{d}\theta_{e}\right)
	\left(\prod_{f\in F}\mathrm{e}^{J_{f}\cos{\Theta^{\prime}_{f}}}\right)\prod_{f\in F
	}\delta(\Theta^{\prime}_{f}-\Theta_{f})\,	\nonumber\\
	&=&\left\{
	\begin{array}{c}
		\left[\frac{1}{\left|\mathrm{det}\mathbf{\tilde{I}}\right|}\int^{\pi}_{-\pi}\left(\prod_{f\in F_{\mathrm{aux}}}\mathrm{d}\Theta^{\prime}_{f}\right)\right]\,\mathcal{Z}=\left[\frac{1}{\left|\mathrm{det}\mathbf{\tilde{I}}\right|} \left(2\pi\right)^{|E|-|F|}\right]\,\mathcal{Z}, \quad |F|\leq|E| \\
		\\
		\left[\frac{1}{\left|\mathrm{det}\mathbf{\tilde{I}}\right|} \left(\frac{\Lambda}{2\pi}\right)^{|F|-|E|}\right]\,\mathcal{Z}, \qquad\qquad\qquad\qquad\qquad\qquad\qquad\, |F|\geq|E|
	\end{array}\right.\nonumber\\
	\nonumber\\
	&=&const.\times \mathcal{Z}\,.
	 \label{proof}
\end{eqnarray}

\subsection{Construction of the Fourier Basis Functions}
\label{ssec:rules}

Here we show how to generate Fourier series expansion (\ref{fourier}) from a $4D$ $U(1)$ LGT, that is, how to generate the basis functions of (\ref{basis}).
In this way a subsystem of our complete model will behave as the target model; more precisely, the Hamiltonian of the complete model on this subsystem will coincide with the (Fourier series of the) Hamiltonian of the target model.
We remark, though, that we will generate the basis functions of (\ref{basis}) only for \emph{bounded}  values of $m_j$, $0 \leq m_j \leq M_j$ for all $j$, where the $M_j$'s are large, positive integers.
We will show this by first introducing the merge and deletion rules, and by then explaining how to obtain arbitrary many--body interactions and build the Fourier basis from these fundamental rules (an explicit construction of many--body interactions with more technical details is given in \ref{construction}).

\begin{figure}[htb]\centering
\psfrag{a}{\small{$\theta_3$}}
\psfrag{b}{\small{$\theta_2$}}
\psfrag{c}{\small{$\theta_1$}}
\psfrag{d}{\small{$\theta_4$}}
\psfrag{e}{\small{$\theta_7$}}
\psfrag{g}{\small{$\theta_6$}}
\psfrag{h}{\small{$\theta_5$}}
\psfrag{1}{\small{$J_{1234}=\infty$}}
\psfrag{2}{\small{$J_{4567}$}}
\psfrag{3}{\small{$J=\infty$}}
\psfrag{4}{\small{$J=0$}}
\psfrag{j}{\small{$J_{134567}$}}
\psfrag{A}{(a)}
\psfrag{B}{(b)}
\psfrag{C}{(c)}
\includegraphics[width=0.8\columnwidth]{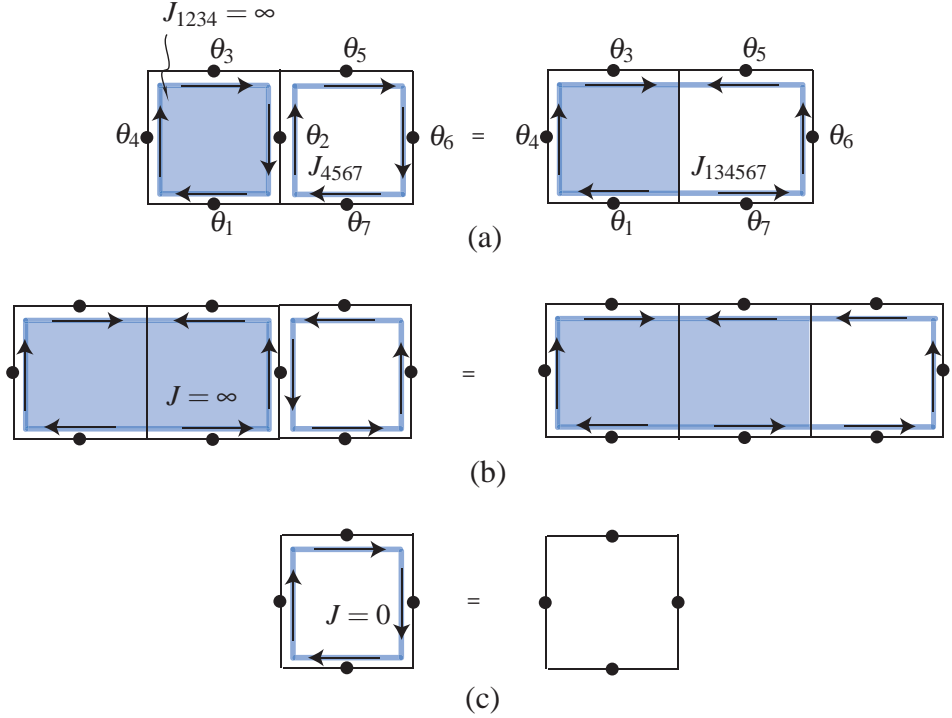}
\caption[]{A blue square represents an interaction in that face, whereas a shaded blue face represents an face with an infinite coupling strength. (a) Merge rule: by letting $J_f\to \infty$ of the left face, this is merged with the face on its right. The resulting face only depends on the spins on its boundary. (b) The resulting face can be merged again with a neighboring face by letting $J_f\to \infty$. Note the resulting direction of the arrows. (c) Deletion rule: (the interaction on) a face is deleted by setting $J_f=0$. }
\label{fig:mergedeletionrule}
\end{figure}

\emph{Merge and deletion rules}.
For simplicity, we consider the state $|\alpha\rangle$ of (\ref{alpha}) defined only on two faces (see Fig.~\ref{fig:mergedeletionrule}(a)):
\begin{eqnarray}
|\alpha_{a,b}\rangle=
\int_{-\pi}^{\pi}
\left(\prod_{e=1}^7 d\theta_e \right) \:
\mathrm{e}^{ J_a \cos\Theta_a} \mathrm{e}^{ J_b \cos \Theta_b}
|\Theta_a\rangle |\Theta_b\rangle\, .
\label{eq:alphaab}
\end{eqnarray}
We define the merge rule on, say, face $a$. Our aim is to obtain a delta function $\delta(\Theta_a)$, in the limit $J_a\to \infty$.
We consider a slight modification of (\ref{eq:alphaab}):
\begin{eqnarray}
|\tilde{\alpha}_{a,b}\rangle =
N(J_a)
\int_{-\pi}^{\pi}
\left(\prod^7_{e=1  , \: e \neq  2} d\theta_e d\Theta_a \right)
\mathrm{e}^{ J_a (\cos\Theta_a-1)}
\mathrm{e}^{ J_b \cos \Theta_b}
|\Theta_a\rangle |\Theta_b\rangle,
\label{eq:alphaab-2}
\end{eqnarray}
where $N(J_a)$ is defined as
\begin{eqnarray}
N(J_a) = \sqrt{\frac{ J_a}{2\pi}}\,.
\end{eqnarray}
This implies that we will obtain the target partition function $\mc{Z}$ with the prefactors:
\begin{eqnarray}
\left(\prod^{M\, \mathrm{terms}}_{m}\frac{\mathrm{e}^{ J_m}}{N(J_m)} \right) \mc{Z} \label{prefactorz}
\end{eqnarray}
where $M$ is the number of faces where the merge rule has been applied, and $J_m$ denotes the coupling strength of the merged face, $J_m\to \infty $. The usual quantity of interest, such as the free energy per particle in the thermodynamic limit, is shifted by a known amount (again, thermodynamic quantities involving derivatives of the free energy are not altered using our regularization method\footnote{The prefactor of the partition function in (\ref{prefactorz}), as well as the additive extra term to the free energy in (\ref{addfactor}), are all formally infinities. As far as these infinities appear in a controllable way, we can get them off by first choosing the coupling strength $J_{m}$'s large but still finite. So that a shift in the free energy has been introduced such as in (\ref{addfactor}). Then we can set $J_{m}$'s to infinity after calculation of observables. No observable will be affected because they all involve a difference or derivative in the free energy. Since the additional term does not contain any physical parameter, it will not contribute even being formally infinite.}):
\begin{eqnarray}
\lim_{N\to \infty}  \frac{- F}{N} &=&
\lim_{N\to \infty} \frac{1}{N}
\ln Z +
\lim_{N\to \infty} \frac{M}{N}
\left(
 \sum_{m}J_m - \frac{1}{2} \ln \frac{ \sum_m J_m}{2\pi} \label{addfactor}
\right)
\end{eqnarray}
where $N$ is the number of particles in the classical $U(1)$ LGT model.

We have
\begin{eqnarray}
N(J_a)
\int_{-\pi}^{\pi}
 d\Theta_a  \mathrm{e}^{ J_a (\cos\Theta_a -1)} = \frac{ \sqrt{2\pi  J_a }}{ \mathrm{e}^{ J_a}} I_{0}( J_a),
\end{eqnarray}
where $I_{0}( J_a)$ is the modified Bessel function of the first kind. The asymptotic behavior of this function is
\begin{eqnarray}
  I_{0}( J_a) \sim
  \frac{\mathrm{e}^{ J_a}}{\sqrt{2\pi  J_a}} ( 1 + \mc{O}(\frac{1}{ J_a}) ), \qquad J_a\to \infty.
\end{eqnarray}
Thus it follows that
\begin{eqnarray}
\lim_{J_a\to \infty} N(J_a) \mathrm{e}^{ J_a (\cos\Theta_a-1)} = \delta(\Theta_a)\, ,
\end{eqnarray}
as desired.
Therefore,
\begin{eqnarray}
&&\lim_{J_a\to \infty}
\int_{-\pi}^{\pi}\left( \prod_{e=1,\: e\neq 2}^7d\theta_e\right)
 N(J_a)
\mathrm{e}^{  J_a (\cos\Theta_a-1)}
\mathrm{e}^{ J_b \cos \Theta_b}
|\Theta_a\rangle |\Theta_b\rangle  \nonumber\\
&=&
\int_{-\pi}^{\pi}d\theta_1   d \theta_3 d \theta_4 d \theta_5 d \theta_6 d \theta_7
\mathrm{e}^{ J_b \cos \tilde{\Theta_b}} |0\rangle |\tilde{\Theta}_b\rangle \, .
\label{eq:mergedface}
\end{eqnarray}
where the variable $\Theta_b$ has become $\tilde{\Theta}_b$ after imposing the constraint of the delta function:
\begin{eqnarray}
\tilde{\Theta}_b = \theta_5 + \theta_ 6 + \theta_ 7 - \theta_ 1 -\theta_ 4 - \theta_ 3\,.
\end{eqnarray}
Note that this corresponds to a $6$--body interaction of the same type, and with interaction strength $J_b$. Thus, letting $J_a\to \infty$ effectively merges face $a$ and $b$. Note also that the variables with opposite sign have opposite pointing directions in the resulting face, see Fig.~\ref{fig:mergedeletionrule}(a).

It is straightforward to see that the same derivation applies for $|\alpha\rangle$ defined on all faces.
In this case, one would only substitute the condition of the delta function, $\Theta_a =0$, on the face with which it has to be merged.
The process can be concatenated, that is, the face resulting from a merge rule can be merged again with a neighboring face, thereby becoming a $8$--body interaction (see Fig. \ref{fig:mergedeletionrule}(b)).
Note that, whenever one face is merged to another, the new edges have the opposite direction than the original ones, as noted above.

The deletion rule works by setting the $J_f=0$, which results in switching off the interaction in that face (see Fig. \ref{fig:mergedeletionrule}(c)).

\emph{Construction of many--body interactions.}
The Fourier basis functions of the set (\ref{basis}) can be generated by making repeated use of the fundamental merge and deletion rules presented above. More specifically, one first generates several ``copies'' of a certain variable, that is, one replicates, say, $m_j$ times the variable $\theta_j$, as required in (\ref{basis}). Then one generates the cosine and the sine of an arbitrary sum of such term (also with the corresponding signs). We refer the reader to \ref{construction} for technical details of these explicit constructions. Here we emphasize that this construction is achieved by applying the merge and the deletion rule on specific faces of the $4D$ lattice, and by fixing some variables using the gauge symmetry of the model (``gauge fixing''). The latter is a procedure that can be carried out so long as the edges whose variable has been fixed do not form a closed loop \cite{Cr77}. This is precisely why we need to resort to a $4D$ lattice, since only then is our construction of the interactions (\ref{basis}) free of closed loops (the same case as in \cite{DDBM1}).

\subsection{The Completeness Result}
\label{completeness}

We are now in the position of proving the main result of this paper.
In $\S\,\ref{quantum}\,$ we have expressed in general the $U(1)$ LGT partition function as an overlap between two quantum states. In $\S\,\ref{ssec:rules}\,$ we have shown that through merge, deletion and gauge fixing in a $4D$ $U(1)$ LGT, any basis functions of a Fourier series expansion can be generated (see also \ref{construction}). As a result, any partition function of the $U(1)$ LGT (with local coupling constants) of the following form can be generated and expressed in terms of a quantum amplitude $\langle\alpha|\psi\rangle$:
\begin{eqnarray}
	&&\mathcal{Z}_{\mathrm{LGT}}\label{LGT}\\
	 &=&\int^{\pi}_{-\pi}\left(\prod_{j}\mathrm{d}\theta_{j}\right)\exp\left\{\sum_{\{m_{j}\}}J^{c}_{m_{j}}\cos\left(\sum_{j}m_{j}\theta_{j}\right)+J^{s}_{m_{j}}\sin\left(\sum_{j}m_{j}\theta_{j}\right)\right\}\nonumber
\end{eqnarray}
with
\begin{eqnarray}
	0 \leq m_{j}\leq M_{j}, \qquad \forall j.\label{range}
\end{eqnarray}
Here all $M_{j}$ are (large) positive integers. Furthermore, in $\S\,\ref{hamiltonian}\,$ we have also expressed a general Hamiltonian with continuous variables in terms of a Fourier series with real coefficients. Let us compare (\ref{LGT}) with the partition function of a general classical model satisfying conditions (\ref{condition1}), (\ref{condition2}) and (\ref{condition3}):
\begin{eqnarray}
	&&\mathcal{Z}_{\mathrm{classical}}\label{classical}\\
	&=&\left(\prod_{j}\int^{b_{j}}_{a_{j}}\mathrm{d}x_{j}\right)\mathrm{e}^{-\beta\mathcal{H}\{x_{j}\}}
	\nonumber\\
	 &=&\mathcal{N}\int^{\pi}_{-\pi}\left(\prod_{j}\mathrm{d}\theta_{j}\right)\nonumber\\&&\cdot\exp\left\{\sum_{\{m_{j}\}}-\beta\Re\mathrm{H}_{\{m_{j}\}}\cos\left(\sum_{j}m_{j}\theta_{j}\right)+\beta\Im\mathrm{H}_{\{m_{j}\}}\sin\left(\sum_{j}m_{j}\theta_{j}\right)\right\}\nonumber
\end{eqnarray}
with $\mathcal{N}=\prod_{j}\frac{b_{j}-a_{j}}{2\pi}$. We realize that if we choose the local coupling constants in (\ref{LGT}) to be equal to the Fourier coefficients (times $\beta$), \textit{i.e.}
\begin{eqnarray}
	 J^{c}_{m_{j}}=-\beta\Re\mathrm{H}_{\{m_{j}\}}, \qquad J^{s}_{m_{j}}=\beta\Im\mathrm{H}_{\{m_{j}\}}, \label{coeff}
\end{eqnarray}
then (up to a constant factor $\mathcal{N}$) the general partition function $\mathcal{Z}_{\mathrm{classical}}$ can be approximated by the partition function $\mathcal{Z}_{\mathrm{LGT}}$. The approximation lies in the fact that in (\ref{classical}) the integers $m_{j}$ run from $0$ to $+\infty$, while in (\ref{LGT}) it runs within a finite range, see (\ref{range}). This approximation can be made to any precision as we increase the repetition of the $\theta_{j}$ variables in the basis functions in (\ref{basis}), see \ref{efficiency}. Therefore, the $4D$ $U(1)$ LGT is complete such that the partition function of any classical model of continuous variables (without constraints) can be mapped to that of the former. That is, by fixing some of the coupling strengthes of a $4D$ $U(1)$ LGT, the remaining subsystem behaves as the continuous model to be simulated.

We emphasize here that models satisfying conditions (\ref{condition1}), (\ref{condition2}) and (\ref{condition3}) are all included in the completeness result, regardless of their dimensions (can be larger, equal to, or smaller than $D=4$) and specific forms of interaction. Also, though the original $4D$ $U(1)$ LGT is gauge invariant (\ref{gaugesymm}), this $U(1)$ gauge symmetry is broken after the mapping because the target model no longer possesses the same gauge symmetry in general.

Finally, we discuss the overhead in the system size of the complete model as a function of the features of the target model. That is, we study how many variables are needed in the complete model in order to generate the Fourier series of (\ref{LGT}) with a truncation at the $M^{\mathrm{th}}$ mode (each $m_{j}\leq M$ in (\ref{range})).
We have seen in \ref{construction} that the generation of each Fourier component (each element of (\ref{basis})) requires a polynomial enlargement of our complete model. \textit{i.e.} In order to generate a single Fourier term such as $\cos(\sum_{j}m_{j}\theta_{j})$, a number of $Poly(\sum_{j}m_{j})$ edge variables are needed, where $Poly(\cdot)$ is a polynomial of its argument. (This fact also implies that the same order $Poly(\sum_{j}m_{j})$ of couplings need to be tuned to infinity or zero in producing this Fourier term. See $\S\,\ref{xymodel}\,$ for a simple and explicit example.)
Thus, the efficiency is measured by the number of Fourier components to be generated in the expansion (\ref{LGT}).
For a single $K$--body interaction term $H^{K}(\{\theta_{j}\})$ in the Hamiltonian (depending on a certain set of $K$ variables $\{\theta_{j}\}$), we need to generate $\sim [Poly(M)]^{K}$ number of Fourier components, where $Poly(M)$ is a polynomial in $M$.
Generally, all combinations of $\theta_{j}$ variables in $K$--body interactions, with $0\leq K\leq N$ ($N$ is the total number of variables), may be present, thus resulting in the scaling
\begin{eqnarray}
	\sim \sum_{K=0}^{N}\left(\begin{array}{c}N \\ K\end{array}\right)[Poly(M)]^{K}\sim\exp{(N)}[Poly(M)]^{K_{\mathrm{max}}}  \,
\end{eqnarray}
for the number of Fourier components.
However, in most cases $K$ does not scale with the system size, $K=K_{\mathrm{max}}$, and moreover only few--body interactions appear in the Hamiltonian (\textit{e.g.} $K=2$ for two body interactions). In this case the scaling is polynomial in both parameters
\begin{eqnarray}
\left(\begin{array}{c}N \\ K_{\mathrm{max}}\end{array}\right)[Poly(M)]^{K_{\mathrm{max}}}\sim Poly(N)[Poly(M)]^{K_{\mathrm{max}}}  \, .
\end{eqnarray}
Note that this question of efficiency in generating Fourier components is unrelated to the question of the accuracy of the approximation of a finite Fourier series (see \ref{efficiency} for the latter).

\section{Examples and Applications of the Completeness Result}
\label{examples}

As an illustration of our completeness result, we give a few example of models whose partition function can be reduced to our class of models.

\subsection{The $XY$ model}
\label{xymodel}

In the $XY$ model we have $2D$ unit vectors (classical spins) $\vec{s}_{i}$ defined on a lattice. The Hamiltonian obeys $O(2)$ (or $U(1)$) symmetry:
\begin{eqnarray}
	 \mathcal{H}_{XY}=-\sum_{<ij>}J_{ij}\ \vec{s}_{i}\cdot\vec{s}_{j}=-\sum_{<ij>}J_{ij}\cos\left(\theta_{i}-\theta_{j}\right), \label{XYhami}
\end{eqnarray}
The partition function of the $XY$ model
\begin{eqnarray}
	 \mathcal{Z}_{XY}=\int^{2\pi}_{0}\left(\prod_{i}\mathrm{d}\theta_{i}\right)\mathrm{e}^{\beta\sum_{<ij>}J_{ij}\cos\left(\theta_{i}-\theta_{j}\right)} \label{XY}
\end{eqnarray}
is seen by itself to be of a $U(1)$ LGT type. Therefore, for this particular case, we do not need to generate the whole Fourier basis, but we just need to generate the $2$--body interactions of (\ref{XYhami}) with the constructions explained in $\S\,\ref{ssec:rules}\,$ (more details in \ref{construction}). Here below we give an explicit pictorial construction of a $1D$ $XY$ model from a $4D$ $U(1)$ LGT as a transparent example of our completeness result. In this particularly simple case we only need a $3D$ sublattice of the $4D$ $U(1)$ LGT for the construction, as shown in Fig. \ref{fig:XY} below. The thick black edges represent variables of the target $XY$ model which eventually will build a $1D$ chain (in terms of interactions). All the odd numbered variables $\theta_{1}$, $\theta_{3}$, $\theta_{5}$, \textit{etc.} are distributed along the same line. Even numbered ones $\theta_{2}$, $\theta_{4}$, \textit{etc.} are distributed separately along the two sides. This arrangement of variables guarantees the correct relative sign in the interaction $\cos(\theta_{i}-\theta_{i+1})$ (to be generated after merge of faces) between every neighboring pair of variables. In the figure it is illustrated how to obtain interactions $\cos(\theta_{1}-\theta_{2})$ and $\cos(\theta_{2}-\theta_{3})$ by merging the blue faces and gauge fixing the red edge variables, and these are the only two typical constructions we need. By direct repetitions all interaction terms of the $XY$ Hamiltonian can be constructed. It is easy to see that in producing an interaction term like $\cos(\theta_{1}-\theta_{2})$, $12$ couplings are taken to the infinity limit; similarly in producing $\cos(\theta_{2}-\theta_{3})$, $8$ couplings are taken to infinity, \textit{etc.}

\begin{figure}[htb]\centering
\psfrag{1}{$\theta_1$}
\psfrag{2}{$\theta_2$}
\psfrag{3}{$\theta_3$}
\psfrag{4}{$\theta_4$}
\psfrag{5}{$\theta_5$}
\psfrag{a}{$J_{12}$}
\psfrag{b}{$J_{23}$}
\psfrag{c}{$J_{34}$}
\psfrag{d}{$J_{45}$}
\includegraphics[width=0.7\columnwidth]{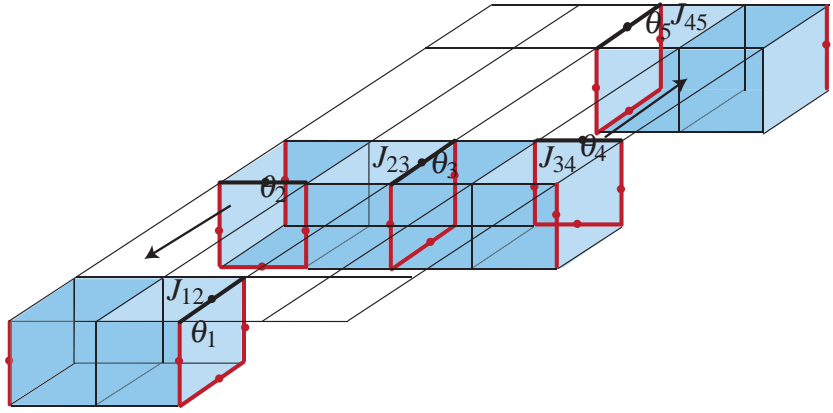}
\caption{Generation of the $1D$ $XY$ model from a $4D$ $U(1)$ LGT. Red edges indicate edges whose variables have been gauge fixed, thick black edges indicate edges containing variables that are present in the target model (they will build the $1D$ chain). Black arrow indicate cubes where the variable is replicated according to Fig. \ref{fig:concatenation-U1} (\ref{construction}). Interactions of the kind $\cos(\theta_i-\theta_{i+1})$ take place in blue prisms, and are a particular case of the interaction shown in Fig. \ref{fig:mdr-difference} (\ref{construction}). Indication of merged faces (blue faces in the figures) has been simplified here to avoid overloading.
}
\label{fig:XY}
\end{figure}

\subsection{The Gaussian and Mean Spherical Models}
\label{sphericalmodel}

In both the Gaussian and the mean spherical models the Hamiltonian (also the `effective' Hamiltonian appearing in the expression of partition functions) is a quadratic form of continuous unbounded spin variables $\sigma_{i}$ ($i=1,2,\ldots,N$):
\begin{eqnarray}
	\mathcal{H}=-\sum_{<ij>}J_{ij}\sigma_{i}\sigma_{j}-\sum_{i}h_{i}\sigma_{i}. \label{hamil}
\end{eqnarray}

In the Gaussian model, each variable $\sigma_{i}$ takes values in $(-\infty, +\infty)$ and is assigned a probability distribution of a Gaussian form:
\begin{eqnarray}
	 p(\sigma_{i})\mathrm{d}\sigma_{i}=\left(\frac{A}{\pi}\right)^{\frac{1}{2}}\mathrm{e}^{-A\sigma^{2}_{i}}\mathrm{d}\sigma_{i}, \qquad i=1, 2, \ldots, N, \label{probability}
\end{eqnarray}
so that the average value is $\langle \sigma^{2}_{i}\rangle=\frac{1}{2A}$. The partition function of the Gaussian model reads
\begin{eqnarray}
	 \mathcal{Z}_{\mathrm{Gauss}}=\left(\frac{A}{\pi}\right)^{\frac{N}{2}}\int^{+\infty}_{-\infty}\left(\prod_{i}\mathrm{d}\sigma_{i}\right)\mathrm{e}^{-A\sum_{i}\sigma^{2}_{i}+\beta\sum_{<ij>}J_{ij}\sigma_{i}\sigma_{j}+\beta\sum_{i}h_{i}\sigma_{i}}. \label{Gauss}
\end{eqnarray}

Alternatively, in the mean spherical model, instead of introducing a probability distribution, we impose a constraint on the average value of $\sum_{i}\sigma^{2}_{i}$ such that
\begin{eqnarray}
	\langle\sum^{N}_{i=1}\sigma^{2}_{i}\rangle=N. \label{average}
\end{eqnarray}
This constraint can be incorporated in the Hamiltonian by introducing a spherical field $\lambda$:
\begin{eqnarray}
	 \mathcal{H}_{\mathrm{mean}}=-\sum_{<ij>}J_{ij}\sigma_{i}\sigma_{j}-\sum_{i}h_{i}\sigma_{i}+\lambda\sum_{i}\sigma^{2}_{i}. \label{varhamil}
\end{eqnarray}
The partition function of the mean spherical model is now
\begin{eqnarray}
	 \mathcal{Z}_{\mathrm{mean}}=\int^{+\infty}_{-\infty}\left(\prod_{i}\mathrm{d}\sigma_{i}\right)\mathrm{e}^{-\beta\lambda\sum_{i}\sigma^{2}_{i}+\beta\sum_{<ij>}J_{ij}\sigma_{i}\sigma_{j}+\beta\sum_{i}h_{i}\sigma_{i}} \label{mean}
\end{eqnarray}
subject to the constraint
\begin{eqnarray}
	 \langle\frac{\partial\mathcal{H}_{\mathrm{mean}}}{\partial\lambda}\rangle=-\frac{1}{\beta}\frac{\partial\ln\mathcal{Z}_{\mathrm{mean}}}{\partial\lambda}=N. \label{lambda}
\end{eqnarray}
Comparing the form (\ref{mean}) with (\ref{Gauss}), we say that the Gaussian model is a mean spherical model with a prescribed spherical field $A$.

For both the Gaussian and the mean spherical models, we can make a cut--off of the $\sigma_{i}$ variables in order to satisfy our condition (\ref{condition1}). That is to say, we assign a joint probability distribution $p\{\sigma_{i}\}$ with a compact support $\mathcal{C}$ to the set of $\sigma_{i}$ variables (any $\sigma_{i}$ vanishes outside some (large) radius in $\mathbb{R}^{N}$):
\begin{eqnarray}
	 p\{\sigma_{i}\}=0, \qquad \forall \sigma_{i}\notin \mathcal{C}, \quad \forall i. \label{support}
\end{eqnarray}
As a result the integrals over $\sigma_{i}$'s in (\ref{Gauss}) and (\ref{mean}) do not extend to infinity:
\begin{eqnarray}
	 \int^{+\infty}_{-\infty}\left(\prod_{i}\mathrm{d}\sigma_{i}\right)\rightarrow\int_{\mathcal{C}}\left(\prod_{i}\mathrm{d}\sigma_{i}\right) \label{integral}
\end{eqnarray}
With this modification of the models, both the Gaussian and the mean spherical models belong to our class which can be approximated by a $4D$ $U(1)$ LGT (the quadratic forms in (\ref{Gauss}) and (\ref{mean}) all have Fourier series expansions).

\subsection{A Model of Coupled Planar Pendulums}
\label{pendulummodel}

\begin{figure}[htb]
	\centering
		\includegraphics[width=0.6\columnwidth]{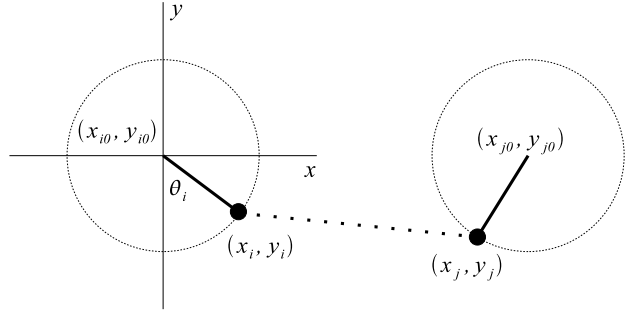}
	\caption{Coupled planar pendulums. The position $(x_{i}, y_{i})$ of each pendulum is determined by the center coordinate $(x_{i0}, y_{i0})$ and an angle $\theta_{i}$. Pendulums are interacting pairwise with the interaction potential depending on the relative distances between their positions.}
	\label{fig:pendulum}
\end{figure}

Consider a number of planar pendulums with pairwise coupling (a discretization of the sine--Gordon model in $1+1$ dimensions), see Fig. \ref{fig:pendulum}.
Let us assume that the interaction potential between any pair of pendulums depends only on their relative distance and each pendulum has a self--interaction depending only on its position (\textit{e.g.} a constant gravitational field). The Hamiltonian can be expressed as
\begin{eqnarray}
	\mathcal{H}=\sum_{i}u_{i}(\cos\theta_{i}, \sin\theta_{i})+\sum_{<ij>}u_{ij}(\cos\theta_{i}, \cos\theta_{j}, \sin\theta_{i}, \sin\theta_{j}, \cos(\theta_{i}-\theta_{j})) \label{potential}
\end{eqnarray}
where in the simplest case the self--interaction
\begin{eqnarray}
	u_{i}\propto J^{x}_{i}\sin\theta_{i}+J^{y}_{i}\cos\theta_{i} \label{ui}
\end{eqnarray}
and the pair interaction
\begin{eqnarray}
	u_{ij}\propto\sqrt{J^{x}_{ij}(\sin\theta_{i}-\sin\theta_{j})+J^{y}_{ij}(\cos\theta_{i}-\cos\theta_{j})
+J_{ij}\cos(\theta_{i}-\theta_{j})} \label{uij}
\end{eqnarray}
with $J_{i}$, $J_{ij}$, \textit{etc}. couplings. It is clear that this Hamiltonian (\ref{potential}) is another example in our class, whose partition function can be mapped to the partition function of the $4D$ $U(1)$ LGT.

\subsection{U(1) LGT with a Monopole Term}
\label{monopole}

Consider the $U(1)$ LGT with a monopole term. This is described by an additional quadratic term supplemented to the standard Wilson action. This system has been studied in connection with finding a confinement mechanism based on monopole condensation \cite{BS,DK,DT}. The action reads
\begin{eqnarray}
	S_{\mathrm{mono}} = J \sum_{a<b,\,n}\left(1-\cos{\Theta_{ab}(n)}\right)+\lambda\sum_{c,\,n}|M_{c}(n)|, \label{monopoleaction}
\end{eqnarray}
where
\begin{eqnarray}
 M_{c}(n)=\frac{1}{4\pi}\epsilon_{cdab}\left(\bar{\Theta}_{ab}(n+d)-\bar{\Theta}_{ab}(n)\right)
\end{eqnarray}
with $\epsilon_{cdab}$ the totally antisymmetric Levi--Civita symbol and $n$ labels the face. The physical flux $\bar{\Theta}_{ab}(n)$ is related to the face variable $\Theta_{ab}(n)$ by
\begin{eqnarray}
 \bar{\Theta}_{ab}(n)=\Theta_{ab}(n)-2\pi \mathcal{N}_{ab}(n)
\end{eqnarray}
with $\mathcal{N}_{ab}(n)$ the number of Dirac strings passing through the face \cite{Dirac}. The last term in (\ref{monopoleaction}) has a Fourier series expansion so that this model is included in the complete $4D$ $U(1)$ LGT.

\subsection{Mean--Field Theory}
\lb{ssec:mft}

Our proof of the main result yields as a by--product the construction of a mean field theory for the $4D$ $U(1)$ LGT. This corresponds to an interaction pattern where all pairwise $4$-body interactions between the edge variables are present. Formally,
the action of the mean field theory reads
\begin{eqnarray}
S_{\mathrm{m.f.}}=\sum_{\forall\, i<j<k<l} J_{ijkl} \cos(\theta_i + \theta_j  + \theta_k +\theta_l).
\end{eqnarray}
Geometrically, all edges are connected with all edges. This action belongs to the class of models whose action we can generate exactly. Therefore our completeness result also includes the mean--field theory for the $U(1)$ LGT.

\section{U(1) Lattice Gauge Theory in a Curved Spacetime}
\label{lgtwithg}

As a further development and also an important application of the completeness result of $\S\,\ref{complete}\,$, we study a $U(1)$ LGT defined in a curved background spacetime and show that it is included in our completeness result. Physically, this situation corresponds to photons in a curved spacetime, as it happens to cosmological photons that have propagated on a curved spacetime during the cosmological evolution. In this section we use the following notation. Latin letters $a$, $b$ \textit{etc.} are used for flat indices and Greek letters $\mu$, $\nu$ \textit{etc.} for curved indices. We also adopt the convention that any minus sign before an index can be dragged out as an overall $(-1)$ factor because it is abelian, \textit{e.g.} $A_{-a}=-A_{a}$. The coupling constant is denoted by $\mathcal{G}$ and the lattice spacing by $\Delta$ as in previous sections.

\subsection{Definition of the Lattice in a Curved Spacetime}

In order to formulate the $U(1)$ LGT in presence of an arbitrary background metric $g_{\mu\nu}$, we first define a \textit{lattice} in a curved spacetime manifold.
In the formulation of the $U(1)$ LGT in $\S\,\ref{basics}\,$ we started with the Euclidean metric $\delta_{ab}=\mathrm{Diag}(1, 1, 1, 1)$ instead of the Minkowski metric $\eta_{ab}=\mathrm{Diag}(-1, 1, 1, 1)$, as the two are related by a Wick rotation. The Wick rotation method has been generalized to curve spacetime \cite{BD}. Given a background spacetime metric, a smooth family of local Wick rotations can be defined on each (co--)tangent vector space of the manifold (thus it is defined on the whole vector bundle). This is done by analytic continuation of either the local time coordinate or the local metric \cite{Liu,More}. Therefore, Lorentzian manifolds can be Wick--rotated to Riemannian manifolds in general. In our formulation below for the curved spacetime, we shall take the metric $g_{\mu\nu}$ to be Riemannian (positive definite) rather than Lorentzian for convenience without loss of generality.

\begin{figure}[htb]
	\centering
		\includegraphics[width=0.5\columnwidth]{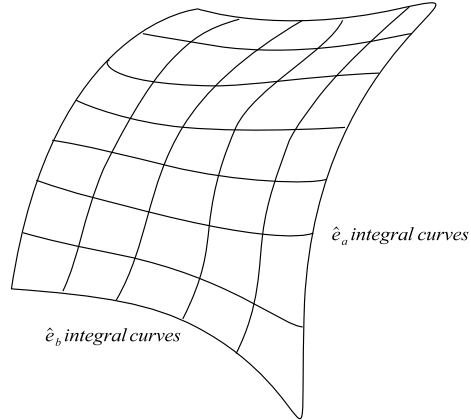}
	\caption{The manifold is covered with a net of intersecting integral curves of the non-coordinate basis vectors. The curves are equally separated so that each small patch (face) is a curved rectangle with equal sides.}
	\label{fig:lattice}
\end{figure}

Now let us take a general Riemannian manifold with a given metric $g_{\mu\nu}$. We could have defined a lattice using the coordinate curves (with respect to the natural coordinates), \textit{i.e.} we pick up a set of hypersurfaces defined by $x^{\mu}=const.$ such that the manifold is filled up by volume cells formed by these hypersurfaces. The intersections of hypersurfaces (coordinate curves) induces a `net' (a graph structure) with `nods' (vertices) linked by segments of coordinate curves (edges). This `net' formed by coordinate curves might give rise to a lattice structure at first sight (once we specify a lattice constant). However, it turns out inconvenient to work with the coordinate basis.\footnote{The natural coordinates may not be orthogonal and the coordinate differences $\Delta x^{\mu}$ and $\Delta x^{\nu}$ ($\mu\neq\nu$) might have different dimensions. It is in general invalid to identify the length of an edge in a lattice with a difference in coordinates such as $\Delta x^{\mu}$.} Therefore, we switch to the non--coordinate basis defined by introducing the frame fields $\mathrm{e}^{a}_{\mu}$ such that
\begin{eqnarray}
	g_{\mu\nu}=\mathrm{e}^{a}_{\mu}\mathrm{e}^{b}_{\nu}\eta_{ab}, \qquad g^{\mu\nu}=\mathrm{e}^{\mu}_{a}\mathrm{e}^{\nu}_{b}\eta^{ab}, \label{vierbein}
\end{eqnarray}
where the fields $\mathrm{e}^{a}_{\mu}$ are also referred to as \textit{tetrad} or \textit{vierbein} fields which bring the metric to a flat one locally. The determinant of the metric field and that of the vierbein are related such that $g\equiv\mathrm{det}[g_{\mu\nu}]=\pm\left(\mathrm{det}[\mathrm{e}^{a}_{\mu}]\right)^{2}$ with the sign determined by the signature of $\eta_{ab}$. Note that in our case we have chosen the flat metric to be Euclidean $\eta_{ab}=\delta_{ab}$. Therefore we have
\begin{eqnarray}
	|\mathrm{det}[\mathrm{e}^{a}_{\mu}]|=\sqrt{\mathrm{det}[g_{\mu\nu}]}\equiv\sqrt{g}. \label{det}
\end{eqnarray}
Formally, we have switched from the general coordinate basis
\begin{eqnarray}
	\left\{\partial_{\mu}\equiv\frac{\partial}{\partial x^{\mu}}\right\} &\quad& \mbox{for the local tangent space},\nonumber\\
	\{dx^{\mu}\} &\quad& \mbox{for the local dual space} \label{coordbasis}
\end{eqnarray}
with metric $g_{\mu\nu}$ to the orthonormal non--coordinate basis
\begin{eqnarray}
	\{\hat{e}_{a}\equiv \mathrm{e}^{\mu}_{a}\partial_{\mu}\} &\quad& \mbox{for the local tangent space},\nonumber\\
	\{\hat{\omega}^{a}\equiv \mathrm{e}^{a}_{\mu}dx^{\mu}\} &\quad& \mbox{for the local dual space} \label{noncoordbasis}
\end{eqnarray}
with a flat metric $\eta_{ab}$. With the introduction of the orthonormal non--coordinate basis, now we can make use of the integral curves of the basis vectors $\{\hat{e}_{a}\}$, $a=0,1,2,3$ in defining a lattice. Note that we do not change the coordination number of the flat or curved lattices. A set of integral curves is orthogonal everywhere so that the `net' formed by these curves defines a curved rectangle (rather than an arbitrarily shaped quadrilateral) on each small `patch' (a face). (See Fig. \ref{fig:lattice}.)

\begin{figure}[htb]
	\centering
		\includegraphics[width=0.6\columnwidth]{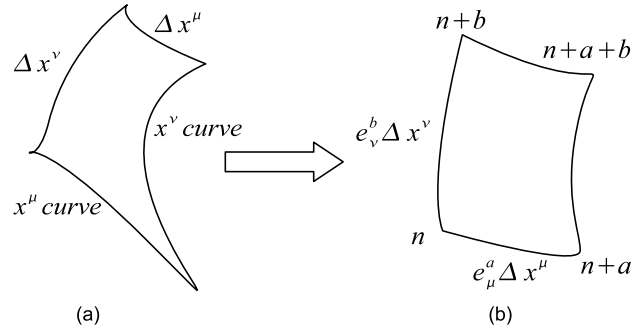}
	\caption{(a) A small patch (a curved quadrilateral) formed by coordinate curves. (b) A curved rectangle with equal sides formed by integral curves of the orthonormal non-coordinate basis.}
	\label{fig:curved face}
\end{figure}

Now we could define a \textit{lattice} by imposing a lattice constant on this `net' formed by integral curves of the non--coordinate basis. It is required that the integral curves are equally separated along each direction $\hat{a}$, $\hat{b}$, \textit{etc} with a common distance $\Delta$. The intersections (`nods' of the `net') of this set of integral curves are taken as vertices, so that every segment of integral curves in between two `nods' becomes an edge. Each edge of the lattice has the same length $\Delta$, see Fig. \ref{fig:curved face}.
In the figure and in the following we shall use the short notation $a=\mathrm{e}^{a}_{\mu}\Delta x^{\mu}$ and $|a|=|b|=\Delta$.

\subsection{The Action of the U(1) LGT in Curved Spacetime}

\begin{figure}[htb]
	\centering
		\includegraphics[width=0.7\columnwidth]{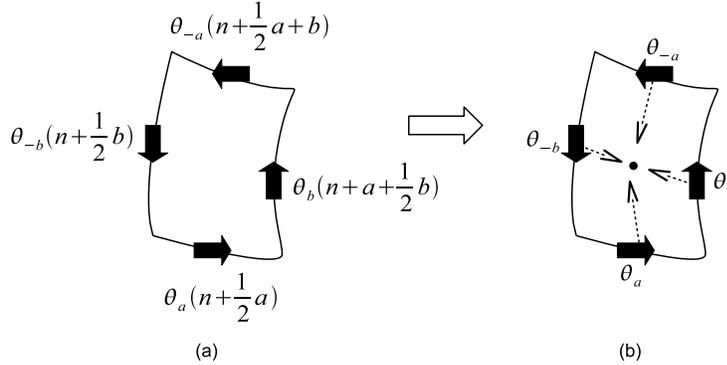}
	\caption{(a) The four edge variables each defined at the center of the corresponding edge. (b) The edge variables have to be parallelly transported to the center (or any other common point) before summing up.}
	\label{fig:curved field}
\end{figure}

With a lattice defined, now we can define the edge variables $\theta_{a}$ in a similar way as in the flat case by
\begin{eqnarray}
	\theta_{a}\equiv\Delta\mathcal{G}A_{a} \label{cedgevar}
\end{eqnarray}
with $\theta_{a}=\mathrm{e}^{\mu}_{a}\theta_{\mu}$ and $A_{a}=\mathrm{e}^{\mu}_{a}A_{\mu}$. Unlike in the flat case, here we have to be careful and specific on the position where exactly the edge variables are defined. It is convenient to put the edge variables in the middle of each edge (see Fig. \ref{fig:curved field}), \textit{e.g.}, take the face with four vertices labeled $n$, $n+a$, $n+b$ and $n+a+b$, the edge variables are written as
\begin{eqnarray}
	\theta_{a}(n+\frac{1}{2}a), \quad \theta_{b}(n+a+\frac{1}{2}b), \quad \theta_{-a}(n+\frac{1}{2}a+b), \quad \theta_{-b}(n+\frac{1}{2}b) \label{cedgevareg}
\end{eqnarray}
on the corresponding edges, respectively.

Next is the face variable $\Theta_{ab}$ which can not be defined by simply taking summation over the four edge variables belonging to the face because those vectors are defined at different tangent spaces (of different points).\footnote{The sum $\sum_{e\in f}\theta_{e}$ in the continuum limit $\Delta\to 0$ approaches $\Delta^{2}\mathcal{G}(F_{ab}+\gamma^{c}_{ab}A_{c})$ instead of the field tensor $F_{ab}$ alone. The extra term involves the structure factor $\gamma^{c}_{ab}$ appearing in the commutation relation of the basis vectors $[\hat{e}_{a}, \hat{e}_{b}]=\gamma^{c}_{ab}\hat{e}_{c}$, which never vanishes identically in a non-coordinate basis unless within a finite globally flat region.} In order to be consistent with the continuum limit, the four edge variables $\theta_{e}$, $e\in f$ for each face have to be parallelly transported to a common point before summation. We choose the center of each face so that the expression of $\Theta_{f}$ can be written in a symmetrical way, see Fig. \ref{fig:curved field}.

We shall denote by $\tilde{\theta}_{e}$ the parallelly transported edge variable $\theta_{e}$ from the center of edge $e$ to the center of face $f$ ($e\in f$). For example, the parallelly transported $\theta_{a}(n+a/2)$ is denoted by $\tilde{\theta}_{a}(n+a/2)$.\footnote{Edge variables are always transported to the center of the face. Therefore for a given face, the location of the transported face is omitted in notation. \textit{e.g.} In $\tilde{\theta}_{a}(n+a/2)$, the argument $n+a/2$ only indicates the location of the edge variable before transport.} Then the face variable is defined by
\begin{eqnarray}
	\Theta_{f}=\sum_{e\in f}\tilde{\theta}_{e}. \label{cfacevar}
\end{eqnarray}
Take a face as the one depicted in Fig. \ref{fig:curved field}.
We have the face variable
\begin{eqnarray}
	 \Theta_{ab}=\tilde{\theta}_{a}(n+\frac{1}{2}a)+\tilde{\theta}_{b}(n+a+\frac{1}{2}b)+\tilde{\theta}_{-a}(n+\frac{1}{2}a+b)+\tilde{\theta}_{-b}(n+\frac{1}{2}b). \label{facevarab}
\end{eqnarray}
The transported $\tilde{\theta}_{e}$ is connected with the original $\theta_{e}$ by the following relation (taking $\theta_{b}(n+b/2)$ transported to $n+a/2+b/2$ for example)
\begin{eqnarray}
	 \tilde{\theta}_{b}(n+\frac{1}{2}b)=\theta_{b}(n+\frac{1}{2}b)+\frac{1}{2}a\,\omega^{c}_{ab}\theta_{c}(n+\frac{1}{2}b), \qquad a\equiv \mathrm{e}^{a}_{\mu}\Delta x^{\mu} \label{para}
\end{eqnarray}
with the spin connection $\omega^{c}_{ab}$ given by
\begin{eqnarray}
	 \omega^{c}_{ab}=\mathrm{e}^{c}_{\nu}\mathrm{e}^{\mu}_{a}(\partial_{\mu}\mathrm{e}^{\nu}_{b}+\mathrm{e}^{\rho}_{b}\Gamma^{\nu}_{\mu\rho}). \label{spin}
\end{eqnarray}
Here $\Gamma^{\rho}_{\mu\nu}$ is the Christoffel connection which is symmetric with respect to the two lower indices, \textit{i.e.} $\Gamma^{\rho}_{\mu\nu}=\Gamma^{\rho}_{\nu\mu}$ (torsion free).
We emphasize two points here:
\begin{enumerate}
	\item In expressions involving a parallel transport such as (\ref{para}), there is no summation over index $a$ if we write $\mathrm{e}^{a}_{\mu}\Delta x_{\mu}$ in replacing $a$  in the last term of the right hand side. It represents a particular direction rather than a dummy index. Therefore we prefer to write directly
\begin{eqnarray}
	\tilde{\theta}_{b}(n+\frac{1}{2}b)=\theta_{b}(n+\frac{1}{2}b)+\Delta\omega^{c}_{ab}\theta_{c}, \qquad |a|\equiv |\mathrm{e}^{a}_{\mu}\Delta x^{\mu}|=\Delta \label{prefer}
\end{eqnarray}
in following parts of the paper (assuming indices $a$, $b$ \textit{etc.} not containing a minus sign for simplicity).
	\item The functions $\omega^{c}_{ab}$ as well as $\theta_{c}$ appearing in the last term are position dependent. They could be defined either at the point $n+b/2$ (as in the expression) or at the point $n+a/2+b/2$. These different choices have the same continuum limit and are equivalent up to $\mathcal{O}(\Delta^{2})$.
\end{enumerate}

Therefore, by plugging (\ref{para}) into (\ref{facevarab}), we obtain
\begin{eqnarray}
	 \Theta_{ab}=&&\theta_{a}(n+\frac{1}{2}a)+\theta_{b}(n+a+\frac{1}{2}b)+\theta_{-a}(n+\frac{1}{2}a+b)+\theta_{-b}(n+\frac{1}{2}b)
	\nonumber\\
	&&-\Delta\omega^{c}_{ab}\theta_{c}+\Delta\omega^{c}_{ba}\theta_{c}. \label{plugin}
\end{eqnarray}
Let us look at the continuum limit. The sum of the first four terms in (\ref{plugin}) gives
\begin{eqnarray}
	\sum_{e\in f}\theta_{e}\stackrel{\Delta\to 0}{\longrightarrow}&&\Delta \mathrm{e}^{\mu}_{a}\partial_{\mu}\theta_{b}-\Delta \mathrm{e}^{\nu}_{b}\partial_{\nu}\theta_{a}
	\nonumber\\
	&=&\Delta \mathrm{e}^{\mu}_{a}\partial_{\mu}(\mathrm{e}^{\nu}_{b}\theta_{\nu})-\Delta \mathrm{e}^{\nu}_{b}\partial_{\nu}(\mathrm{e}^{\mu}_{a}\theta_{\mu})
	\nonumber\\
	&=&\Delta \mathrm{e}^{\mu}_{a}\mathrm{e}^{\nu}_{b}(\partial_{\mu}\theta_{\nu}-\partial_{\nu}\theta_{\mu})+\Delta (\mathrm{e}^{\mu}_{a}\partial_{\mu}\mathrm{e}^{\nu}_{b}-\mathrm{e}^{\mu}_{b}\partial_{\mu}\mathrm{e}^{\nu}_{a})\theta_{\nu} . \label{limitsum}
\end{eqnarray}
On the other hand, using (\ref{spin}), we see that the last two terms in (\ref{plugin}) equal
\begin{eqnarray}
	&&-\Delta(\omega^{c}_{ab}-\omega^{c}_{ba})\theta_{c}
	\nonumber\\
	 &=&-\Delta\left(\mathrm{e}^{\mu}_{a}(\partial_{\mu}\mathrm{e}^{\nu}_{b}+\mathrm{e}^{\lambda}_{b}\Gamma^{\nu}_{\mu\lambda})-\mathrm{e}^{\mu}_{b}(\partial_{\mu}\mathrm{e}^{\nu}_{a}+\mathrm{e}^{\lambda}_{a}\Gamma^{\nu}_{\mu\lambda})\right)\mathrm{e}^{c}_{\nu}\theta_{c}
	\nonumber\\
	 &=&-\Delta(\mathrm{e}^{\mu}_{a}\partial_{\mu}\mathrm{e}^{\nu}_{b}-\mathrm{e}^{\mu}_{b}\partial_{\mu}\mathrm{e}^{\nu}_{a})\theta_{\nu}-\Delta(\mathrm{e}^{\mu}_{a}\mathrm{e}^{\lambda}_{b}\Gamma^{\nu}_{\mu\lambda}-\mathrm{e}^{\mu}_{b}\mathrm{e}^{\lambda}_{a}\Gamma^{\nu}_{\mu\lambda})\theta_{\nu}
	\nonumber\\
	 &=&-\Delta(\mathrm{e}^{\mu}_{a}\partial_{\mu}\mathrm{e}^{\nu}_{b}-\mathrm{e}^{\mu}_{b}\partial_{\mu}\mathrm{e}^{\nu}_{a})\theta_{\nu}.
	\label{limitterm}
\end{eqnarray}
In obtaining (\ref{limitterm}) we have used the symmetry properties of the connection $\Gamma^{\rho}_{\mu\nu}$ with respect to the two lower indices. As a consequence of (\ref{limitsum}) and (\ref{limitterm}), the face variable in the continuum limit approaches
\begin{eqnarray}
	\Theta_{ab}&=&\sum_{e\in f}\theta_{e}-\Delta(\omega^{c}_{ab}-\omega^{c}_{ba})\theta_{c}
	\nonumber\\
	&\stackrel{\Delta\to 0}{\longrightarrow}&\Delta \mathrm{e}^{\mu}_{a}\mathrm{e}^{\nu}_{b}(\partial_{\mu}\theta_{\nu}-\partial_{\nu}\theta_{\mu})
	\nonumber\\
	&=&\Delta^{2}\mathcal{G} \mathrm{e}^{\mu}_{a}\mathrm{e}^{\nu}_{b}(\partial_{\mu}A_{\nu}-\partial_{\nu}A_{\mu})
	\nonumber\\
	&=&\Delta^{2}\mathcal{G} \mathrm{e}^{\mu}_{a}\mathrm{e}^{\nu}_{b}F_{\mu\nu}=\Delta^{2}\mathcal{G} F_{ab}. \label{limitface}
\end{eqnarray}

Therefore, with the face variable defined in (\ref{cfacevar}) yielding a proper continuum limit (\ref{limitface}), we defined the action of the $U(1)$ LGT with a background metric as follows:
\begin{eqnarray}
	S=\frac{1}{2\mathcal{G}^{2}}\sum_{f\in F}[1-\cos\Theta_{f}]. \label{caction}
\end{eqnarray}
This expression resembles the action in (\ref{action}) for the flat metric case (with different definitions of the faces and face variables). We shall prove that (\ref{caction}) leads to the correct continuum limit. As $\Delta\to 0$, we have each $\Theta_{f}$ being small, so that
\begin{eqnarray}
	S\stackrel{\Delta\to 0}{\longrightarrow}\frac{1}{2\mathcal{G}^{2}}\sum_{f\in F}\frac{1}{2}\Theta^{2}_{f}
		 =\frac{1}{2\mathcal{G}^{2}}\sum_{ab}\frac{1}{2}\Delta^{4}\mathcal{G}^{2}F^{2}_{ab}=\frac{1}{4}\sum_{ab}\Delta^{4}F^{2}_{ab}. \label{limitaction}
\end{eqnarray}
Furthermore, we have
\begin{eqnarray}
	\Delta^{4}&=&|(\mathrm{e}^{a}_{\mu}\Delta x^{\mu}){}_{\wedge}(\mathrm{e}^{b}_{\nu}\Delta x^{\nu}){}_{\wedge}(\mathrm{e}^{c}_{\rho}\Delta x^{\rho}){}_{\wedge}(\mathrm{e}^{d}_{\sigma}\Delta x^{\sigma})|
	\nonumber\\
	&\stackrel{\Delta\to 0}{\longrightarrow}&|\hat{\omega}^{a}{}_{\wedge}\hat{\omega}^{b}{}_{\wedge}\hat{\omega}^{c}{}_{\wedge}\hat{\omega}^{d}|
	=|\mathrm{det}[\mathrm{e}^{a}_{\mu}]|\mathrm{d}^{4}x=\sqrt{g}\,\mathrm{d}^{4}x \label{invvolume}
\end{eqnarray}
being the invariant volume. In obtaining (\ref{invvolume}) we have used (\ref{det}) with $\wedge$ the wedge (exterior) product and $\hat{\omega}^{a}$ \textit{etc.} the dual basis. On the other hand,
\begin{eqnarray}
	 F^{2}_{ab}&=&F_{ab}F_{cd}\eta^{ac}\eta^{bd}=F_{ab}F_{cd}\mathrm{e}^{a}_{\mu}\mathrm{e}^{c}_{\rho}g^{\mu\rho}\mathrm{e}^{b}_{\nu}\mathrm{e}^{d}_{\sigma}g^{\nu\sigma}
	\nonumber\\
	&=&F_{\mu\nu}F_{\rho\sigma}g^{\mu\rho}g^{\nu\sigma}=F_{\mu\nu}F^{\mu\nu}. \label{contra}
\end{eqnarray}
As a result, by combining (\ref{invvolume}) and (\ref{contra}), we prove that
\begin{eqnarray}
	S\stackrel{\Delta\to 0}{\longrightarrow}\frac{1}{4}\int\mathrm{d}^{4}x\sqrt{g}\,F_{\mu\nu}F^{\mu\nu}. \label{prooflimit}
\end{eqnarray}
That is, the discrete $U(1)$ LGT action defined in (\ref{caction}) approaches the pure $U(1)$ gauge action with a background metric in the continuum limit.\footnote{We should have instead of the field tensor $F_{\mu\nu}=\partial_{\mu}A_{\nu}-\partial_{\nu}A_{\mu}$, the covariant field tensor $\mathcal{F}_{\mu\nu}=D_{\mu}A_{\nu}-D_{\nu}A_{\mu}$ with the covariant derivative $D_{\mu}A_{\nu}=\partial_{\mu}A_{\nu}-\Gamma^{\rho}_{\mu\nu}A_{\rho}$ appearing in (\ref{proof}). However, since the connection $\Gamma^{\rho}_{\mu\nu}$ is symmetric with respect to the two lower indices while the field tensor is antisymmetric, we have $F_{\mu\nu}=\mathcal{F}_{\mu\nu}$ identically.}

The expression of the action defined in (\ref{caction}) with the face variable given in (\ref{cfacevar}) involves explicitly the lattice constant $\Delta$ because of (\ref{prefer}) and (\ref{plugin}). It is possible to rewrite the explicit expression of the face variable $\Theta_{ab}$ (\ref{plugin}) in a more similar way to that in the flat case (\ref{Theta}). Let us look at the last two terms in (\ref{plugin}). As shown in (\ref{limitterm}), the Christoffel connections (see (\ref{spin})) cancel out; each remaining term depends on the lattice constant and a derivative of the vierbein field in such a way that it is identified as a first order \textit{differential} of the vierbein field. Therefore it can be rewritten as a \textit{difference} in the vierbein up to the order $\mathcal{O}(\Delta^{2})$, \textit{e.g.}
\begin{eqnarray}
	\Delta \mathrm{e}^{\mu}_{a}(\partial_{\mu}\mathrm{e}^{\nu}_{b})\theta_{\nu}\sim [\mathrm{e}^{\nu}_{b}(n+a)-\mathrm{e}^{\nu}_{b}(n)]\theta_{\nu}. \label{diff}
\end{eqnarray}
Therefore, the last two terms in (\ref{plugin}) can be split into four terms (to be more symmetric) each involving a difference of two vierbeins, one defined at the center of the face, the other at one edge. \textit{i.e.}
\begin{eqnarray}
	&&-\Delta(\omega^{c}_{ab}\theta_{c}-\omega^{c}_{ba})\theta_{c}
	 =-\Delta(\mathrm{e}^{\mu}_{a}\partial_{\mu}\mathrm{e}^{\nu}_{b}-\mathrm{e}^{\mu}_{b}\partial_{\mu}\mathrm{e}^{\nu}_{a})\theta_{\nu}
	\nonumber\\
	&\sim&\quad [\mathrm{e}^{\mu}_{a}(n+\frac{1}{2}a+\frac{1}{2}b)-\mathrm{e}^{\mu}_{a}(n+\frac{1}{2}a)]\theta_{\mu}(n+\frac{1}{2}a)
	\nonumber\\
	 &&\,\,+[\mathrm{e}^{\mu}_{b}(n+\frac{1}{2}a+\frac{1}{2}b)-\mathrm{e}^{\mu}_{b}(n+a+\frac{1}{2}b)]\theta_{\mu}(n+a+\frac{1}{2}b)
	\nonumber\\
	 &&\,\,-[\mathrm{e}^{\mu}_{a}(n+\frac{1}{2}a+\frac{1}{2}b)-\mathrm{e}^{\mu}_{a}(n+\frac{1}{2}a+b)]\theta_{\mu}(n+\frac{1}{2}a+b)
	\nonumber\\
	 &&\,\,-[\mathrm{e}^{\mu}_{b}(n+\frac{1}{2}a+\frac{1}{2}b)-\mathrm{e}^{\mu}_{b}(n+\frac{1}{2}b)]\theta_{\mu}(n+\frac{1}{2}b). \label{term}
\end{eqnarray}
Now with this new form (\ref{term}), the face variable can be rewritten in the following form:
\begin{eqnarray}
	\Theta_{f}\equiv\Theta_{ab}&=&\sum_{e\in f}\theta_{e}-\Delta(\omega^{c}_{ab}-\omega^{c}_{ba})\theta_{c}
	\nonumber\\
	&\sim&\quad  \theta_{\mu}(n+\frac{1}{2}a)\mathrm{e}^{\mu}_{a}(n+\frac{1}{2}a+\frac{1}{2}b)
	\nonumber\\
	&&\,\,+\theta_{\mu}(n+a+\frac{1}{2}b)\mathrm{e}^{\mu}_{b}(n+\frac{1}{2}a+\frac{1}{2}b)
	\nonumber\\
	&&\,\,-\theta_{\mu}(n+\frac{1}{2}a+b)\mathrm{e}^{\mu}_{a}(n+\frac{1}{2}a+\frac{1}{2}b)
	\nonumber\\
	&&\,\,-\theta_{\mu}(n+\frac{1}{2}b)\mathrm{e}^{\mu}_{b}(n+\frac{1}{2}a+\frac{1}{2}b)
	\nonumber\\
	&=&\sum_{e\in f}\theta_{\mathrm{edge}}\cdot \mathrm{e}_{\mathrm{face}}.
		\label{finalface}
\end{eqnarray}
In the last line of (\ref{finalface}) we have used a concise notation to indicate that the face variable $\Theta_{f}$ is a sum over the edge variables $\theta_{e}$ with each edge variable coupled to the vierbein $\mathrm{e}^{\mu}_{a}$ at the center of the face. The form (\ref{finalface}) resembles (\ref{Theta}) in the flat case. We also realize that in this form the vierbein fields need only to be defined for each face. Moreover, by rearranging terms, this form can also be written in the following way
\begin{eqnarray}
	\Theta_{ab}&=&\quad  [\theta_{\mu}(n+\frac{1}{2}a)-\theta_{\mu}(n+\frac{1}{2}a+b)]\mathrm{e}^{\mu}_{a}(n+\frac{1}{2}a+\frac{1}{2}b)
	\nonumber\\
	 &&\,\,+[\theta_{\mu}(n+a+\frac{1}{2}b)-\theta_{\mu}(n+\frac{1}{2}b)]\mathrm{e}^{\mu}_{b}(n+\frac{1}{2}a+\frac{1}{2}b).
		\label{alter}
\end{eqnarray}
The expression (\ref{alter}) is transparent in taking the continuum limit (reducing to $F_{ab}$) as well as the `flat' limit ($\mathrm{e}^{\mu}_{a}=\delta^{\mu}_{a}$).

We emphasize here that a consistent theory of the $U(1)$ LGT with a background metric is not unique. (There are two requirements only: $1)$ The recovery of a $U(1)$ pure gauge theory with a metric in the continuum limit; $2)$ The recovery of a flat $U(1)$ LGT in the limit $g_{\mu\nu}=\eta_{\mu\nu}$, and these do not defined the theory uniquely.) We are making a particular choice so that the resulting theory takes a form similar to the flat theory and the form of the face variable looks more symmetric with respect to the edge variables involved.\footnote{Had we made another choice, \textit{e.g.} had we preferred the edge variable $\theta_{\nu}$ in (\ref{diff}) and $\theta_{\mu}$ in (\ref{term}) be defined at the center of the face rather than on the edge, the face variable could have taken a different form such as
\begin{eqnarray}
	\Theta_{ab}&=& [\theta_{\mu}(n+\frac{1}{2}a)-\theta_{\mu}(n+\frac{1}{2}a+\frac{1}{2}b)]\mathrm{e}^{\mu}_{a}(n+\frac{1}{2}a)
	\nonumber\\
	 &+&[\theta_{\mu}(n+a+\frac{1}{2}b)-\theta_{\mu}(n+\frac{1}{2}a+\frac{1}{2}b)]\mathrm{e}^{\mu}_{b}(n+a+\frac{1}{2}b)
	\nonumber\\
	 &-&[\theta_{\mu}(n+\frac{1}{2}a+b)-\theta_{\mu}(n+\frac{1}{2}a+\frac{1}{2}b)]\mathrm{e}^{\mu}_{b}(n+\frac{1}{2}a+b)
	\nonumber\\
	 &-&[\theta_{\mu}(n+\frac{1}{2}b)-\theta_{\mu}(n+\frac{1}{2}a+\frac{1}{2}b)]\mathrm{e}^{\mu}_{b}(n+\frac{1}{2}b).
		\nonumber
\end{eqnarray}
Also, had we parallelly transported the four edge variables to a common point other than the center of the face, the face variable $\Theta_{ab}$ might have taken a less symmetric form.}

\subsection{The Partition Function of the `Curved' U(1) LGT and the Completeness Result}

We have defined the action of the $U(1)$ LGT coupled to a background gravitational field (a background metric). The action is given by (\ref{caction}) which takes a similar form as in the flat case (\ref{action}). The face variables are defined in (\ref{cfacevar}) (by summing over edge variables parallelly transported to the center of each face) or in (\ref{finalface}) equivalently (by summing over edge variables coupled with the vierbein field at the center of each face). As a result, the partition function of the model reads
\begin{eqnarray}
\mathcal{Z}_{\mathrm{curved}}&=&\int\left(\prod_{e\in E}\mathrm{d}\theta_{e}\right)\exp\left\{-\frac{1}{2\mathcal{G}^{2}}\sum_{f\in F}\left[1-\cos{\left(\sum_{e\in f}\tilde{\theta}_{e}\right)}\right]\right\}
\nonumber\\
&=&\int\left(\prod_{e\in E}\mathrm{d}\theta_{\mathrm{edge}}\right)\exp\left\{-\frac{1}{2\mathcal{G}^{2}}\sum_{f\in F}\left[1-\cos{\left(\sum_{e\in f}\theta_{\mathrm{edge}}\cdot \mathrm{e}_{\mathrm{face}}\right)}\right]\right\}.\nonumber\\ \label{cpartition}
\end{eqnarray}
The action (\ref{caction}) of the $U(1)$ LGT with a curved background metric satisfies the three conditions (\ref{condition1}), (\ref{condition2}) and (\ref{condition3}) in $\S\,\ref{hamiltonian}\,$. Therefore, we conclude that our completeness result contains this model, \textit{i.e.}, that the $U(1)$ LGT with a curved background metric can be mapped to a $4D$ $U(1)$ LGT with a flat metric.

\section{Generalizations}
\label{generalize}

The completeness result obtained in $\S\,\ref{completeness}\,$ can be generalized to include an even broader class of classical models. The generalization is achieved by relaxing the third condition (\ref{condition3}) in $\S\,\ref{hamiltonian}\,$. In other words, we shall consider classical physical systems with dynamical variables subject to constraints in this section. A very important set of physical models are classical Heisenberg models. For these models the spin variables $s$ are classical but subject to the constraint $s_1^2 + s_2^2 + ...+ s_n^2 = 1$, \textit{i.e.} they live in an internal space which is an $S^{n-1}$ sphere.

\subsection{Models with Constraints}
\label{genecons}

Let us take a classical system satisfying only conditions (\ref{condition1}) and (\ref{condition2}). Assume that there are a set of $M$ ($M<N$) independent \textit{unsolvable} constraints\footnote{If a constraint is solvable, then we could release the constraint by a change of variables such that there will be no constraint over this new set of independent variables.} over the variable $x_{j}$'s such that
\begin{eqnarray}
	\mathcal{F}_{l}\{x_{j}\}=0, \qquad l=1,2,\ldots,M. \label{constraints}
\end{eqnarray}
The functions $\mathcal{F}_{l}$ are also assumed to have Fourier series expansions. After normalization of the variables $x_{j}\to\theta_{j}$, we write the Hamiltonian as $\mathcal{H}\{\theta_{j}\}$ and the constraints as $\mathcal{F}\{\theta_{j}\}$. So that the partition function reads (see (\ref{classical}))
\begin{eqnarray}
	 \mathcal{Z}_{\mathrm{classical}}=\mathcal{N}\int^{\pi}_{-\pi}\left(\prod_{j}\mathrm{d}\theta_{j}\right)\left(\prod_{l}\delta\left(\mathcal{F}_{l}\{\theta_{j}\}\right)\right)\mathrm{e}^{-\beta\mathcal{H}\{\theta_{j}\}}. \label{zdelta}
\end{eqnarray}
Now for each constraint $\mathcal{F}_{l}$ we associate an additional variable $\kappa_{l}$ and re-write each delta function as a Fourier transform. Then (\ref{zdelta}) becomes
\begin{eqnarray}
	 \mathcal{Z}_{\mathrm{classical}}&=&\mathcal{N}\int^{\pi}_{-\pi}\left(\prod_{j}\mathrm{d}\theta_{j}\right)\int^{+\infty}_{-\infty}\left(\prod_{l}\frac{\mathrm{d}\kappa_{l}}{2\pi}\right)\nonumber\\&&\cdot\exp\left\{-\beta\mathcal{H}\{\theta_{j}\}+\mathrm{i}\sum_{l}\kappa_{l}\mathcal{F}_{l}\{\theta_{j}\}\right\}. \label{ztransf}
\end{eqnarray}
If we make a (large ultraviolet) cut--off on each of the $\kappa_{l}$ variables such that $\kappa_{l}\in [-\Lambda_{l},+\Lambda_{l}]$, then the integrals $\int^{+\infty}_{-\infty}$ become $\int^{+\Lambda_{l}}_{-\Lambda_{l}}$. Now we can normalize these variables by $\phi_{l}=\frac{\pi}{\Lambda_{l}}\kappa_{l}$. Then the partition function (\ref{ztransf}) takes the following form approximately
\begin{eqnarray}
	 \mathcal{Z}_{\mathrm{classical}}\sim\mathcal{NN'}\int^{\pi}_{-\pi}\left(\prod_{j,l}\mathrm{d}\theta_{j}\mathrm{d}\phi_{l}\right)\mathrm{e}^{\mathcal{H}\{\theta_{j},\phi_{l}\}} \label{zphi}
\end{eqnarray}
with $\mathcal{N'}=\prod_{l}\frac{\Lambda_{l}}{2\pi^{2}}$ and an effective complex Hamiltonian $\mathcal{H}\{\theta_{j},\phi_{l}\}$ defined by
\begin{eqnarray}
	 \mathcal{H}\{\theta_{j},\phi_{l}\}=-\beta\mathcal{H}\{\theta_{j}\}+\mathrm{i}\sum_{l}\frac{\Lambda_{l}}{\pi}\phi_{l}\mathcal{F}_{l}\{\theta_{j}\} \label{effect}
\end{eqnarray}
This Hamiltonian in (\ref{effect}) with additional variables $\phi_{l}$ has a Fourier series expansion
\begin{eqnarray}
	 \mathcal{H}\{\theta_{j},\phi_{l}\}=\sum_{\{m_{j},m_{l}\}}&&\mathrm{H}_{\{m_{j},m_{l}\}}\cos\left(\sum_{j,l}m_{j}\theta_{j}+m_{l}\phi_{l}\right)
	\nonumber\\
	&+&\tilde{\mathrm{H}}_{\{m_{j},m_{l}\}}\sin\left(\sum_{j,l}m_{j}\theta_{j}+m_{l}\phi_{l}\right) \label{complex}
\end{eqnarray}
with complex Fourier coefficients $\mathrm{H}_{\{m_{j},m_{l}\}}$ and $\tilde{\mathrm{H}}_{\{m_{j},m_{l}\}}$. Therefore, if we allow for complex coupling constants $J_{\{m_{j},m_{l}\}}=\mathrm{H}_{\{m_{j},m_{l}\}}$ and $\tilde{J}_{\{m_{j},m_{l}\}}=\tilde{\mathrm{H}}_{\{m_{j},m_{l}\}}$, the partition function $\mathcal{Z}_{\mathrm{classical}}$ in (\ref{zphi}) is of a $U(1)$ LGT type. As a result, if we assume that the cut--off (a regularization) in the delta function representation above works universally, then the $4D$ $U(1)$ LGT with general complex coupling constants is a complete model for classical statistical models with continuous variables subject to constraints.


\subsection{Models with discrete variables}
\label{discrete}

Another possible generalization is to consider classical models with discrete degrees of freedom, such as the Ising model, the Potts model, the vertex models, and the $\mathbb{Z}_{q}$ LGT where the dynamical variables take values only in a discrete set. These models can be considered a special case of continuous models with constraints.

Let us again consider a classical system satisfying only conditions (\ref{condition1}) and (\ref{condition2}). Each of the variables $x_{j}$ can assume only finite discrete values
\begin{eqnarray}
	x_{j}=X_{l_{j}}, \qquad a_{j}\leq X_{l_{j}}\leq b_{j}, \qquad l_{j}=1,2,\ldots,L_{j}. \label{discon}
\end{eqnarray}
The possible values $X_{l_{j}}$ and the number of possible values $L_{j}$ may depend on $j$. It is equivalent to express these constraints by the following expression with $\delta$-functions
\begin{eqnarray}
	\prod_{j}\left[\sum^{L_{j}}_{l_{j}=1}\delta\left(x_{j}-X_{l_{j}}\right)\right] \label{delcon}
\end{eqnarray}
such that for an arbitrary function of a set of $x_{j}$'s, we have
\begin{eqnarray}
	 \left(\prod_{j}\int^{b_{j}}_{a_{j}}\mathrm{d}x_{j}\right)\prod_{j}\left[\sum^{L_{j}}_{l_{j}=1}\delta\left(x_{j}-X_{l_{j}}\right)\right]f\{x_{j}\}=\sum^{L_{N}}_{l_{N}=1}\cdots\sum^{L_{1}}_{l_{1}=1}f\{x_{j}=X_{l_{j}}\}. \label{inttosum}
\end{eqnarray}
In the last summation of (\ref{inttosum}) we are summing over all possible configurations (all possible combination of $x_{j}$ values). Moreover, after normalization of the variables $x_{j}\to\theta_{j}$, $\mathcal{H}\{x_{j}\}\to\mathcal{H}\{\theta_{j}\}$, the delta functions become
\begin{eqnarray}
\delta\left(x_{j}-X_{l_{j}}\right)\to\frac{2\pi}{b_{j}-a_{j}}\delta\left(\theta_{j}-\frac{2\pi}{b_{j}-a_{j}}\left(X_{l_{j}}-\frac{a_{j}+b_{j}}{2}\right)\right). \label{deltran}
\end{eqnarray}
We define
\begin{eqnarray}
	\Theta_{l_{j}}=\frac{2\pi}{b_{j}-a_{j}}\left(X_{l_{j}}-\frac{a_{j}+b_{j}}{2}\right), \qquad \forall\ l_{j}, j \label{thetalj}
\end{eqnarray}
for notational convenience. Consequently, the partition function of a classical model with discrete degrees of freedom can be expressed as
\begin{eqnarray}
	\mathcal{Z}_{\mathrm{discrete}}&=&\sum_{\mathrm{all}\ \{l_{j}\}}\mathrm{e}^{-\beta\mathcal{H}\{x_{j}=X_{l_{j}}\}} \label{pardis}\\
	 &=&\int^{\pi}_{-\pi}\left(\prod_{j}\mathrm{d}\theta_{j}\right)\prod_{j}\left[\sum_{l_{j}}\delta\left(\theta_{j}-\Theta_{l_{j}}\right)\right]\mathrm{e}^{-\beta\mathcal{H}\{\theta_{j}\}}.\nonumber
\end{eqnarray}
Now, with these particular forms of $\delta$-functions appearing in (\ref{pardis}), it is not necessary to go to the region of a complex Hamiltonian. Let us represent each $\delta$-function by a limit
\begin{eqnarray}
	\delta\left(\theta_{j}-\Theta_{l_{j}}\right) = \lim_{\epsilon_{j}\to 0}\delta_{\epsilon_{j}}\left(\theta_{j}-\Theta_{l_{j}}\right)\label{delrep}
\end{eqnarray}
with each $\delta_{\epsilon_{j}}\left(\theta_{j}-\Theta_{l_{j}}\right)$ being a positive--definite function on $[-\pi,\pi]$ depending on the parameter $\epsilon_{j}$. (This function could be a Gaussian or a `square' function.) If we make a cut--off approximation on these $\epsilon_{j}$ parameters, then $\mathcal{Z}_{\mathrm{discrete}}$ can be rewritten as
\begin{eqnarray}
	&&\mathcal{Z}_{\mathrm{discrete}}\{\epsilon_{j}\}\label{cut}\\
	 &\sim&\frac{1}{\left(2\pi\right)^{N}}\int^{\pi}_{-\pi}\left(\prod_{j}\mathrm{d}\theta_{j}\right)\exp\left\{-\beta\mathcal{H}\{\theta_{j}\}+\sum_{j}\ln\left[\sum_{l_{j}}\delta_{\epsilon_{j}}\left(\theta_{j}-\Theta_{l_{j}}\right)\right]\right\}.\nonumber
\end{eqnarray}
The expression on the exponential of (\ref{cut}) can be interpreted as an effective (real) Hamiltonian which allows a Fourier series expansion over $\theta_{j}$'s with real coefficients. Therefore we conclude that $U(1)$ LGT is complete also for certain classical discrete models, again assuming that the cut--off (regularization) (\ref{delrep}) works universally.

\section{Conclusion}
\label{conclusion}

We have proven that any classical partition function depending on continuous variables subject to conditions (\ref{condition1}), (\ref{condition2}) and (\ref{condition3}) can be approximated (to an arbitrary precision) by the partition function of a $4D$ $U(1)$ LGT. In the proof we first introduced a quantum representation of the $U(1)$ LGT partition function. Then through merging and deletion of gauge field variables and proper choices of local coupling constants, a mapping from a $4D$ $U(1)$ LGT partition function to a more general partition function is established. In this sense the $4D$ $U(1)$ LGT is a complete model for a large class of classical models. The completeness result is also generalized to include continuous models with constraints (if we are allowing complex coupling constants) and discrete models.
As a further development and important application of the completeness result, we have developed a consistent theory of the $U(1)$ LGT coupled to a background metric.
The action is defined in a form (\ref{caction}) very close to that of the model in flat spacetime.
Our completeness result holds for this model such that its partition function can be mapped (approximated to an arbitrary precision) to the partition function of a $U(1)$ LGT in flat spacetime. This is the first time that a completeness result is established for continuous statistical models.

We believe that our completeness result cannot be proven with a $3D$ $U(1)$ LGT. Another open question is whether a similar result can be found for non--Abelian LGTs (e.g. for $SU(2)$ or $SU(3)$ LGTs). We envisage that these theories may require a new approach since a direct generalization of our construction seems not to be possible.

\ack
The authors would like to thank Maarten Van den Nest for fruitful discussions and valuable suggestions. The work
is supported by the FWF (SFB-F40) and the European Union (QICS, NAMEQUAM), the Spanish MICINN grant FIS2009-10061, CAM research consortium QUITEMAD S2009-ESP-1594, European Commission PICC: FP7 2007-2013. grant no. 249958, UCM-BS grant GICC-910758.

\appendix

\section{Construction of Many--Body Interactions}
\label{construction}

To generate the functions of the set (\ref{basis}) we will proceed similarly as in \cite{DDBM1,DDBM2}, namely we will make use of the merge and the deletion rule, and of the gauge fixing of edges.
The latter is a procedure for fixing the values of the variables in the lattice, which results in a theory that is physically equivalent to the original one. This can be carried out by virtue of the gauge symmetry of the model. The only restriction in this procedure is that the edges whose variables are fixed by the gauge cannot form a closed loop \cite{Cr77}.

\begin{figure}[htb]\centering
\psfrag{a}{$\theta_1$}
\psfrag{b}{$\theta_2$}
\psfrag{c}{$\theta_3$}
\psfrag{r}{$\alpha_1$}
\psfrag{s}{$\alpha_2$}
\includegraphics[width=0.5\columnwidth]{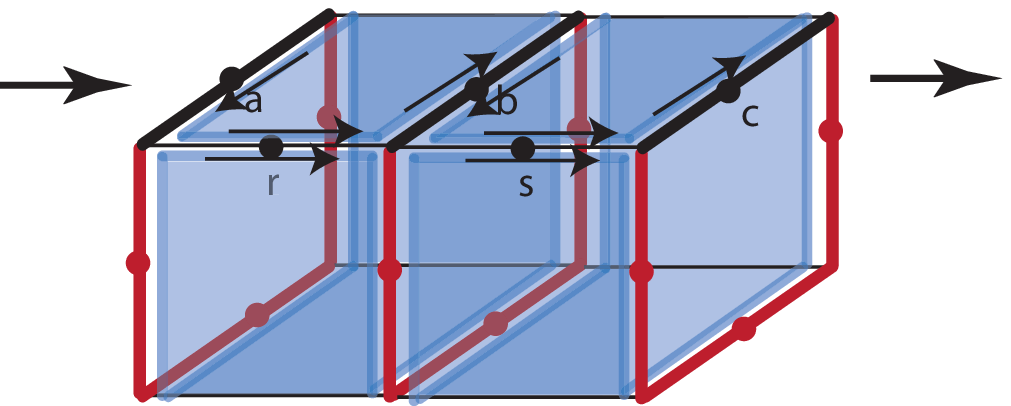}
\caption{Propagation of the variable $\theta_1$ across the $4D$ square lattice (the figure shows only a $3D$ projection of this space).
In all figures, red edges denote edges whose variable has been fixed by the gauge.
By means of the merge rule, $\theta_1$ is propagated into $\theta_3$, since $\theta_2 = -\theta _1$ and $\theta_3= -\theta_2$.
 }
\label{fig:concatenation-U1}
\end{figure}

To construct the many--body interactions, we first need to ``propagate'' the variables inside the lattice in order to bring them close together to interact. This propagation is achieved with the following construction (see Fig.~\ref{fig:concatenation-U1}). On a cube (on the left of Fig.~\ref{fig:concatenation-U1}), we merge the face at the front, at the bottom, and at the back to generate the interaction $J_f\cos(\alpha_1-\alpha_1 + \theta_1+\theta_2)$. Then, we let this coupling strength go to infinity, $J_f\to\infty$, which imposes the constraint $\theta_1 + \theta_2 = 0$. The same process is repeated for the cube on the right of Fig.~\ref{fig:concatenation-U1}, where the constraint reads $\theta_2+\theta_3=0$. Thus, we have set $\theta_3 = \theta_1$, that is, $\theta_1$ has ``propagated'' two sites to the right.
Note that if $\theta_1$ were propagated an odd number of times, the resulting variable would equal $-\theta_1$. This can be circumvented by letting $\theta_1$ participate in the final interaction with the opposite sign (as explained below).
In order to see how to turn the propagation path, we refer the reader to the explanations on Fig. 11 of \cite{DDBM2}, since the construction is analogous.

\begin{figure}[htb]\centering
\psfrag{a}{$\theta_1$}
\psfrag{b}{$\theta_2$}
\psfrag{c}{$\theta_3$}
\psfrag{r}{$\alpha_1$}
\psfrag{s}{$\alpha_2$}
\psfrag{1}{$\theta_5$}
\psfrag{2}{$\theta_4$}
\psfrag{3}{$\theta_6$}
\psfrag{u}{$\alpha_4$}
\psfrag{w}{$\alpha_5$}
\psfrag{j}{$\alpha_3$}
\includegraphics[width=0.5\columnwidth]{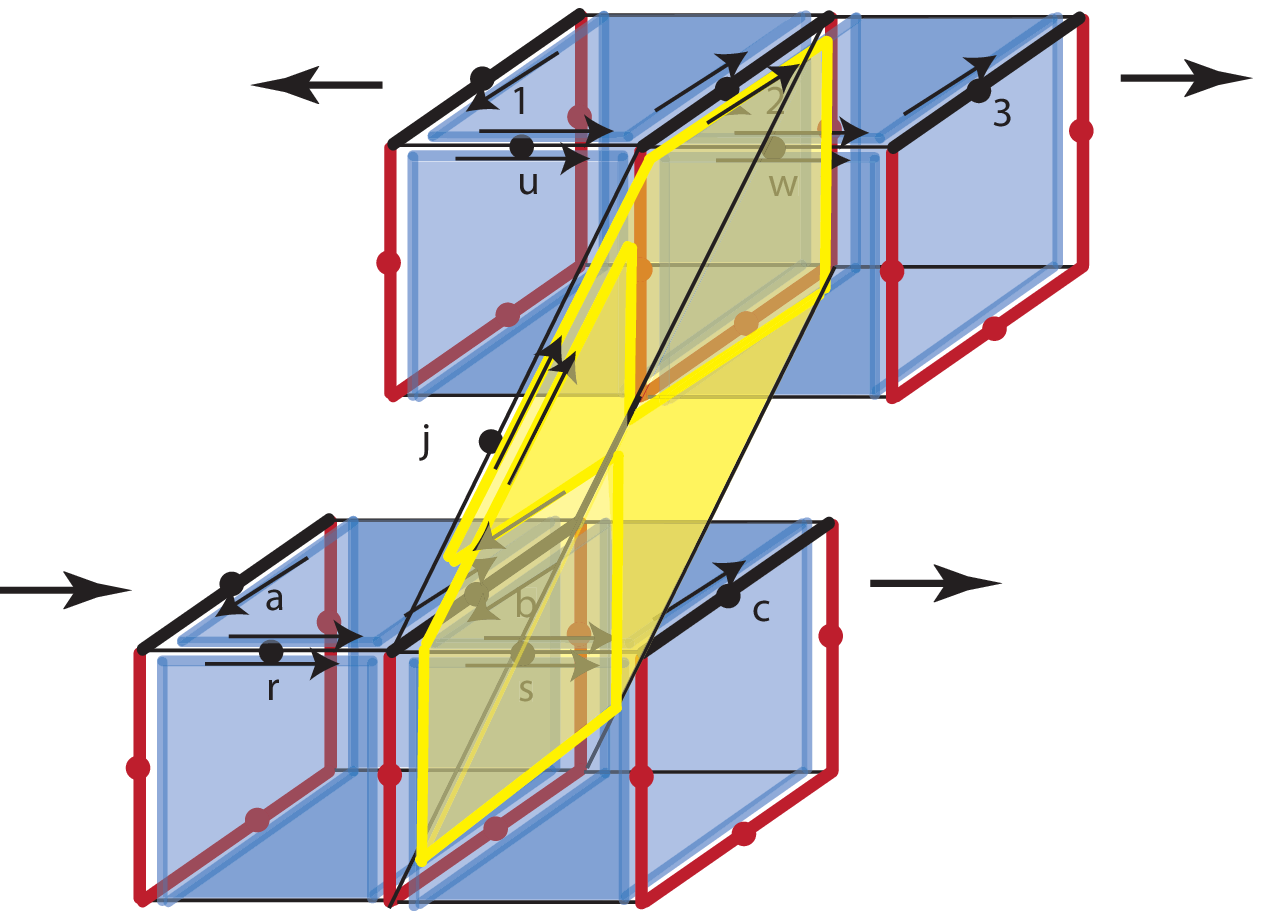}
\caption{Yellow faces are in the direction of the fourth dimension and  have the same meaning as blue faces (\textit{i.e.} merged faces). Replication of the variable $\theta_1$ into $\theta_3$, $\theta_5$ and $\theta_6$. This replication is essentially a propagation (as the one of Fig.~\ref{fig:concatenation-U1}) in the fourth dimension, i.e.~the variable lives now in another $3D$ space.}
\label{fig:4D-replication-U1}
\end{figure}

Now we focus on the replication of the (classical, continuous) variables, that is, on the generation of several copies of a given variable. This is achieved by applying the propagation procedure explained above into the fourth dimension, as shown in Fig.~\ref{fig:4D-replication-U1}. The interaction in the yellow cube is of the form $J_f\cos(\theta_2+\alpha_3+\theta_4 -\alpha_3)$, on which we let  $J_f\to \infty$ and thereby impose $\theta_2 = -\theta_4$ (i.e.~we apply the merge rule on this cube as well). The rest works exactly as the propagation explained in Fig.~\ref{fig:concatenation-U1}.
We note that the reason for using a fourth dimension in the replication of edge variables is that all schemes we have found in three dimensions involve closed loops of variables fixed by the gauge \cite{DDBM1}.

\begin{figure}[htb]\centering
\psfrag{b}{$\theta_1$}
\psfrag{d}{$\theta_2$}
\psfrag{a}{$\theta_3$}
\psfrag{c}{$\theta_4$}
\psfrag{e}{$\theta_5$}
\psfrag{t}{$\alpha_1$}
\psfrag{p}{$\alpha_2$}
\psfrag{s}{$\alpha_3$}
\psfrag{t}{$\alpha_4$}
\psfrag{u}{$\alpha_5$}
\psfrag{r}{$\alpha_6$}
\psfrag{w}{$\alpha_7$}
\psfrag{f}{$J_f$}
\psfrag{A}{$\sigma$}
\includegraphics[width=0.6\columnwidth]{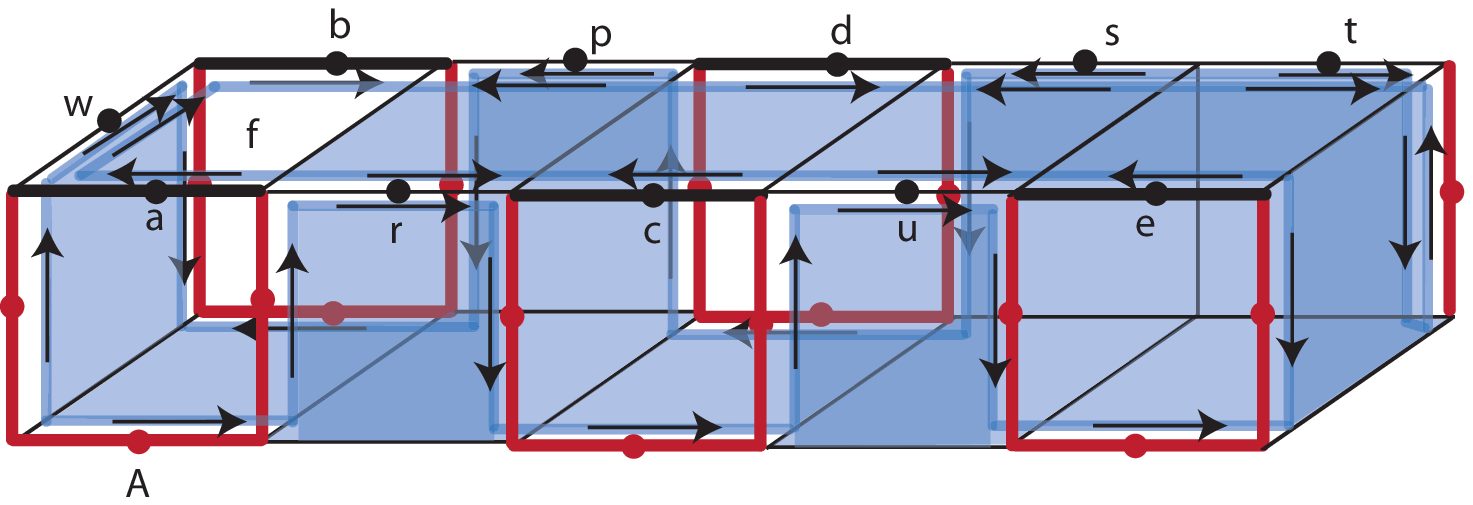}
\caption{
Bold, black edges contain variables that participate in the final interaction $\theta_1,\ldots, \theta_5$.
Red edges (except for the one marked with $\sigma)$ stand for edges whose variables have been fixed to zero using the gauge symmetry, and blue faces stand for merged faces, as in Fig.~\ref{fig:mergedeletionrule}.
If $\sigma =0$, as the other spins,
the large blue face contains the five--body interaction $J_f \cos(\theta_1 +\theta_2+\theta_3+\theta_4+\theta_5 )$, whereas if $\sigma=-\pi/2$, it depends on $J_f\sin(\theta_1 +\theta_2+\theta_3+\theta_4+\theta_5 )$.}
\label{fig:mdr-sum}
\end{figure}

Next we show how to generate interactions of the type (\ref{basis}).
The generation of the interactions $\cos(\sum_{i=1}^K\theta_i)$  for the specific case $K=5$ is illustrated in Fig.~\ref{fig:mdr-sum}. We first propagate each of the variables $\theta_1, \ldots, \theta_5$ inside the lattice in order to distribute them on the edges of the rectangular prism shown in Fig.~\ref{fig:mdr-sum}. Then we merge all faces on the exterior surface of this prism into one large, blue face. As shown above, this face only depends on the spins at its boundaries, that is, it has the form
\begin{eqnarray}
J_f \cos(&&\theta_1 - \alpha_2 +\theta_2 -\alpha_3 +\alpha_4 -\alpha_4  +\alpha_3 + \alpha_2-\alpha_7 \nonumber\\&&+\alpha_6 + \alpha_5 + \theta_5 - \alpha_5 + \theta_4 -\alpha_6 +\theta_3 +\alpha_7)\, .
\end{eqnarray}
Since the dependence on every auxiliary variable $\alpha_i$ cancels, it takes the desired form $J_f \cos(\sum_{i=1}^5 \theta_i)$.
The generalization to any $K$ is straightforward.
For odd $K$, we construct a longer or shorter prism than that of Fig.~\ref{fig:mdr-sum}, arranging $(K-1)/2$ variables in front of the remaining $(K-1)/2$ (as $\theta_1$ and $\theta_2$ are arranged in front of $\theta_3$ and $\theta_4$). The remaining variable would be unpaired, and the vertical edge on the corner of the prism fixed by the gauge (just as $\theta_5$).
For even $K$ the construction is simpler. For $K=4$ we arrange the spins $\theta_1,\ldots,\theta_4$ as in Fig.~\ref{fig:mdr-sum}, and we merge the large, blue face over the face joining the red u--shapes of $\theta_2$ and $\theta_4$ on the right. For any other even $K$, we similarly arrange $K/2$ variables in front of the remaining ones.
Finally, note that the generation of interactions of the type $\cos(2\theta_1)$ is achieved by first replicating the variable $\theta_1$, and then letting its two copies participate in a two--body interaction as explained above.

\begin{figure}[htb]\centering
\psfrag{a}{$\theta_1$}
\psfrag{b}{$\theta_2$}
\psfrag{c}{$\theta_3$}
\psfrag{d}{$\theta_4$}
\psfrag{e}{$\theta_5$}
\psfrag{r}{$\alpha_1$}
\psfrag{w}{$\alpha_2$}
\psfrag{u}{$\alpha_3$}
\psfrag{t}{$\alpha_4$}
\psfrag{s}{$\alpha_5$}
\psfrag{p}{$\alpha_6$}
\psfrag{q}{$\alpha_7$}
\psfrag{v}{$\alpha_8$}
\psfrag{f}{$J_f$}
\includegraphics[width=0.6\columnwidth]{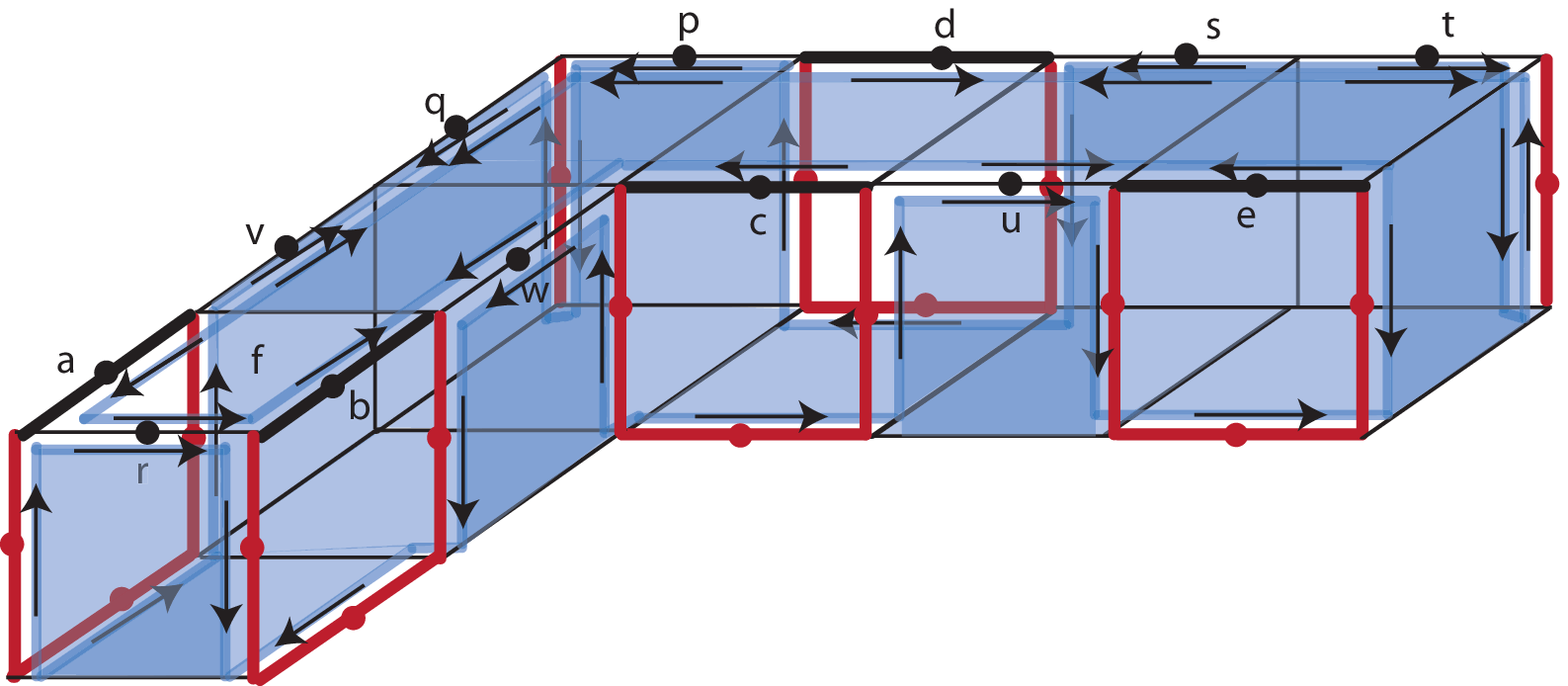}
\caption[]{Five--body interaction $J_f \cos(\theta_1 +\theta_2-\theta_3-\theta_4-\theta_5 )$. The meaning of the symbols is the same as in Fig.~\ref{fig:mdr-sum}.}
\label{fig:mdr-difference}
\end{figure}

Now we only need to show how to generate interactions involving variables with different signs. The generation of $J_f\cos(\theta_1 + \theta_2+\theta_3-\theta_4 -\theta_5)$ is shown in Fig.~\ref{fig:mdr-difference}. The variables which have the same relative sign are arranged as explained for the cosine of the sum of variables (see Fig.~\ref{fig:mdr-sum}). The new element here is that the two sets of variables which have the opposite relative sign must arranged perpendicularly to each other. Then a large, blue face is merged over the external faces, where the desired interaction takes place. One can also verify that the dependency on the auxiliary variables $\alpha_i$ cancels out, and that variables perpendicular to each other have the opposite sign.
The generalization to an interaction $J_f\cos(\sum_{i=1}^{K_1} \theta_i - \sum_{j=1}^{K_2} \theta_j)$ is also straightforward. One only has to arrange the first set of variables as explained for the case $J_f\cos(\sum_{i=1}^{K_1} \theta_i )$, with $K_1$ either odd or even. The other set of variables is arranged also as explained above (with $K_2$ being odd or even), and set perpendicular to the first set.

Finally, we point out that the sine functions are generated by making use of the relation $\sin(\gamma) = \cos (\gamma - \pi/2)$. This phase amounts to gauge fixing one of the spins to $-\pi/2$ instead of 0, as the rest. For example, to generate $\sin(\theta_1+\theta_2+\theta_3+\theta_4+\theta_5)$ we construct the interaction of Fig.~\ref{fig:mdr-sum}, and we fix $\sigma = -\pi/2$.

This shows how one can generate all interactions of the set (\ref{basis}) starting from a $4D$ $U(1)$ LGT.
The construction also shows that each Fourier basis function (i.e.~each term in the set (\ref{basis})) requires a polynomial enlargement in the number of variables of the $4D$ $U(1)$ LGT. We shall return to this fact in $\S\,\ref{completeness}\,$, where we specify the overhead in the system size of the complete model as a function of the features of the target model.

\section{Accuracy of the Finite Fourier Series}
\label{efficiency}

In the completeness result of $\S\,\ref{completeness}\,$, we have made a truncation in the Fourier series basis (see (\ref{LGT}) and (\ref{range})) so as to approximate a general Hamiltonian as an expansion. Here comes a question of accuracy. That is, given a truncation $-M\leq m_{j}\leq M,\, \forall j$ in the Fourier modes,
how close is the following finite Fourier series
\begin{eqnarray}
	 \mathcal{F}_{M}\left[H^{(K)}\right]\equiv\sum_{\{m_{j}\}}H^{(K)}_{\{m_{j}\}}\exp{\left\{\mathrm{i}\sum_{j}m_{j}\theta_{j}\right\}}
\end{eqnarray}
to the original Hamiltonian function $H^{(K)}({\{\theta_{j}\}})$ of a $K$-body interaction term?
According to \cite{Schlag}, for a smooth enough (usually at least differentiable) single variable function $f(\theta)\in C^{\alpha}[-\pi,\pi]$ with $\alpha>0$, we have
\begin{eqnarray}
	|\mathcal{F}_{M}\left[f\right]-f(\theta)|\leq A(f)\sum_{P(\alpha)}\frac{1}{M^{P(\alpha)}}
\end{eqnarray}
where $\mathcal{F}_{M}\left[f\right]$ is a finite Fourier series of $f(\theta)$ with $2M+1$ terms (truncated at the $M^{\mathrm{th}}$ mode as in our case), $A(f)$ is a finite factor depending on the function form of $f(\theta)$ on the domain only, and the finite sum is over some polynomials $P(\alpha)$ of $\alpha$. Therefore we see that in order to have an accuracy $\sim \frac{1}{\mathcal{N}}$ for the finite Fourier series of a single variable, we need a polynomially large truncation $Poly(\mathcal{N})$ in the Fourier modes. We could say that we have a polynomial accuracy in this case. Now with a $K$-body interaction term which is a function of $K$ variables, in order to have an accuracy
\begin{eqnarray}
	|\mathcal{F}_{\mathcal{M}}\left[H^{(K)}\right]-H^{(K)}({\{\theta_{j}\}})|\leq\frac{1}{\mathcal{N}},
\end{eqnarray}
we shall need $\mathcal{M}\sim [Poly(\mathcal{N})]^{K}$ terms in the finite Fourier series. \textit{i.e.} the $K^{\mathrm{th}}$ power of some polynomial of $\mathcal{N}$. Therefore it is still efficient if the model has only few--body interactions (\textit{e.g.} nearest neighbor interactions only) or if $K$ scales polynomially with the system size.


\section*{References}



\end{document}